\documentclass[a4paper,11pt]{article}

\pdfoutput=1

\usepackage[utf8]{inputenc}
\usepackage{dsfont}
\usepackage{epsfig}
\usepackage{slashed}
\usepackage{bbold}
\usepackage{psfrag}
\usepackage{color}
\PassOptionsToPackage{caption=false}{subfig}
\usepackage{subfig}
\usepackage{multirow}
\usepackage{booktabs}
\usepackage{pstricks}
\usepackage{feynmp}
\usepackage{jheppub} 
\usepackage[normalem]{ulem}

\newcommand{\units}[1]{\textrm{\ #1}}

\usepackage{colortbl}

\definecolor{LightCyan}{rgb}{0.88,1,1}
\definecolor{piggypink}{rgb}{0.99, 0.87, 0.9}
\definecolor{applegreen}{rgb}{0.55, 0.71, 0.0}
\definecolor{darkpastelgreen}{rgb}{0.01, 0.75, 0.24}
\definecolor{green-yellow}{rgb}{0.68, 1.0, 0.18}
\definecolor{nicered}{rgb}{0.7,0.1,0.1}
\definecolor{nicegreen}{rgb}{0.1,0.5,0.1}
\definecolor{darkblue}{rgb}{0.0,0.0,.4}
\definecolor{darkred}{rgb}{0.4,0.0,0.0}

\definecolor{red}{rgb}{1.0, 0, 0}

\def\lsim{\mathrel{\rlap{\lower4pt\hbox{\hskip1pt$\sim$}}
     \raise1pt\hbox{$<$}}}         
\def\gsim{\mathrel{\rlap{\lower4pt\hbox{\hskip1pt$\sim$}}
     \raise1pt\hbox{$>$}}}

\newcommand{\mpl}{M_{\rm Pl}}

\def\beqn{\begin{eqnarray}} 
\def\eeqn{\end{eqnarray}} 
\def\be{\begin{equation}}
\def\ee{\end{equation}}

\title{\begin{center}
Crunching Away the Cosmological Constant Problem: \\Dynamical Selection of a Small $\Lambda$\end{center}}

\author[a]{Itay M. Bloch,}
\affiliation[a]{School of Physics and Astronomy, Tel-Aviv University, Tel-Aviv 69978, Israel}
\emailAdd{itay.bloch.m@gmail.com}

\author[b]{Csaba Cs\'aki,}
\affiliation[b]{Department of Physics, LEPP, Cornell University, Ithaca, NY 14853, USA}
\emailAdd{csaki@cornell.edu}

\author[a]{Michael Geller,}
\emailAdd{mic.geller@gmail.com}

\author[a]{Tomer Volansky}
\emailAdd{tomerv@post.tau.ac.il}

\abstract
{We propose a novel  explanation for the smallness of 
the observed cosmological constant (CC).  Regions of space with a large CC are short lived and are dynamically driven to  crunch soon after the end of inflation.  Conversely, regions with a small CC 
 are metastable and long lived and are the only ones to survive until late times.   
While the mechanism assumes many domains  with different CC values,  it does not result in eternal inflation nor does it require a long period of inflation to populate them.
  We present a concrete dynamical model, based on a super-cooled first order phase transition in a hidden conformal sector, that may successfully implement such a crunching mechanism. We find that the mechanism can only solve the CC problem up to the weak scale, above which new physics, such as supersymmetry,  is needed to solve the CC problem all the way to the UV cutoff scale. The absence of experimental evidence for such new physics already implies a mild little hierarchy problem for the CC.   Curiously, in this approach the weak scale arises as the geometric mean of the temperature in our universe today and the Planck scale, hinting on a new ``CC miracle", motivating new physics at the weak scale independent of electroweak physics.  We further predict the presence of new relativistic degrees of freedom in the CFT that should be visible in the next round of CMB  experiments.   Our mechanism is therefore experimentally falsifiable and predictive.
}

\begin{document} 

\maketitle

\section{Introduction} 
\label{sec:intro}

The smallness of the measured Cosmological Constant\footnote{Here and below we denote the CC using a dimension-1 parameter $\Lambda$, while its observed value is $\Lambda_{\rm obs}$.} (CC), $\Lambda_{\rm obs}^4 \simeq 26\units{meV}^4$,  is considered to pose one of the deepest conundrums (for reviews, see e.g.~\cite{Weinberg:1988cp,Martin:2012bt}) in physics. In fact the smallness of the CC introduces three intriguing questions, with (widely) varying levels of severity:
\begin{enumerate}
\item {\bf Why is $\Lambda_{\rm obs}^4$ so small?}  Indeed, its measured value is roughly 120 orders of magnitude below the naive estimate of $\Lambda^4 \simeq M_{\rm Pl}^4$, with $M_{\rm Pl}\simeq 1.2\times 10^{19}~{\rm GeV}$ the Planck scale, and  roughly 48 orders of magnitude below the QCD scale.  In the absence of a dynamical explanation, the CC must be severely fine tuned to agree with the observed value. This is known as {\it the  CC problem}~\cite{Zeldovich:1968ehl}. 
\item {\bf Why is $\Omega_{\Lambda,0} \simeq 2\,  \Omega_{m,0}$?}  A priori, one expects  the  measured CC energy density, $\Omega_{\Lambda,0}$, and the matter energy density today, $\Omega_{m,0}$, to be unrelated.  Consequently, the ratio of these two quantities could have been many orders of magnitude above or below one, while in practice it is measured to be close to unity. This is the so called {\it coincidence problem}~\cite{Fitch:1997cf}.  Within the standard cosmological model, the above implies that the CC energy density comes to dominate the universe only at very low redshift, 
and hence this problem is often referred to as the {\it why now?} problem.
\item {\bf Is the CC related to the TeV scale?}  Numerically, one finds that $\Lambda_{\rm obs}\simeq {\rm TeV}^2/\bar M_{\rm Pl}$ with $\bar M_{\rm Pl} \simeq 2.4\times 10^{18}\units{GeV}$ the reduced Planck scale.   While this relation could be nothing more than a numerical coincidence, it may be viewed as an intriguing hint towards a possible relation between the solution to the CC problem and new physics at the weak scale.
 \end{enumerate}
 In this paper we discuss a solution which naturally addresses all of the above conundrums at once.

Several approaches for explaining the smallness of $\Lambda_{\rm obs}^4$ have been considered (for reviews, see e.g.~\cite{Weinberg:1988cp, Nobbenhuis:2004wn,Polchinski:2006gy,Bousso:2007gp,Yoo2012,Padilla2015,Copeland2006,Sola:2013gha}).  
For example, one approach is to use  some (softly broken) symmetry to set the CC close to zero (see e.g.~\cite{Zumino:1974bg,Bellazzini:2013fga,Coradeschi:2013gda}; for an interesting variation on this idea see~\cite{Witten:1995rz}). This approach has so far been unsuccessful due to the absence of a consistent low-energy symmetry that can sufficiently suppress the value of the  CC. Another approach is to dynamically relax the CC, driving the system toward a small value (see e.g.~\cite{Abbott:1984qf,Steinhardt:2006bf}).  
However, the Weinberg no-go theorem~\cite{Weinberg:1988cp} -- stating that one cannot find a coupled scalar-gravity system with a classical solution in which the CC vanishes without significant tuning -- 
 severely restricts the possible adjustment mechanisms and solutions which circumvent it often result in an empty universe.
 Other approaches include some form of violation of the equivalence principle, such as non-local corrections to gravity (see e.g.~\cite{ArkaniHamed:2002fu}), 
alluding to  thermodynamics and the holographic principle (for a review see~\cite{Banks:2018jqo}), use of graviton compositeness~\cite{Sundrum:1997js}, sequestering of vacuum energy~\cite{Kaloper:2013zca}, unimodular gravity~\cite{Anderson:1971pn}, and even more exotic attempts (see e.g.~\cite{Linde:1988ws}).

Many more attempts were made over the years to solve this problem, and yet perhaps the only wildly acceptable solution to the CC, originally introduced by Weinberg,  invokes the  anthropic principle~\cite{Weinberg:1987dv,Weinberg:1988cp}: living observers should only exist in a universe which allows for structure to form and life to develop.   Weinberg argued that if many regions of space (a ``multiverse") with distinct values of the CC exist,  observers would only be expected to live in domains which exhibit a sufficiently small CC.  
The anthropic solution requires a theory that allows for a ``landscape" of vacua as well as a dynamical mechanism to populate the multiverse. Since the CC in a given patch starts to dominate after a while, eternal inflation seems inevitable.  
Therefore, the solution suffers from two shortcomings.  First, by construction there are no observable experimental consequences for the presence of the multiverse. Second, eternal inflation~\cite{Vilenkin:1983xq,Gibbons:1984hx} has an inherent measure problem (see e.g.~\cite{Linde:1993xx,Linde:2010xz,Freivogel2011} 
) of calculating probabilities across infinite causally disconnected regions, crucially impeding  its predictive power.

In this paper we propose a new direction to address the CC problem.  We assume the population of many 
regions of space, each with a different CC, similarly to the standard anthropic approach. However, in sharp contrast with this approach, we also add dynamics that render all regions metastable.  Any region which goes through the phase transition to the true vacuum suffers a drop in its energy density down to a large and negative value, causing it to  collapse into a big-crunch singularity~\cite{Coleman:1980aw}.  The temperature-dependent decay rate
depends on the cosmological evolution
which in turn depends on the value of the CC in that region.  Large CC comes to dominate the energy density early, triggering a secondary phase of inflation and rapidly driving the region down to the nucleation temperature, below which the phase transition proceeds efficiently.   Conversely, in regions with smaller CC values, the temperature drops slowly as the universe expands and thus those regions live longer.  Only  domains with a small CC survive until today, which explains why we have observed such a small value of the CC around us.

Since all the domains crunch, the secondary phase of inflation is only temporary, thereby evading eternal inflation and its accompanied measure problem. Moreover, as we shall see, a long period of inflation is unnecessary for the population of the landscape, and only a small number of e-folds are required. Meanwhile, the Weinberg no-go theorem is not violated as no special deference is given in the Lagrangian to small values of the CC.   

The solution not only addresses the CC problem but also the coincidence problem.  Indeed, had the CC dominated the energy density in our universe much earlier, it would have driven our Hubble patch to an early  secondary phase of inflation followed up by a phase transition and subsequent crunch long before today.   Smaller values of the CC are possible, however those are presumably  sparse in the multiverse.  

Remarkably, this dynamical solution is falsifiable, with two major phenomenological implications. Firstly, we find  a limit on the maximal CC in the landscape, beyond which we cannot apply the crunching mechanism. The scale of this CC turns out to be (for a more accurate estimate see Sec.~\ref{sec:CCmax}),
\begin{equation}
\Lambda_{\rm max} \sim {\cal O} (\sqrt{T_{0} \bar M_{Pl}}) \simeq {\cal O} ({\rm TeV})\,,
\end{equation}
suggesting that the TeV scale should emerge as a special scale, above which the CC would have to be canceled by some other mechanism, for example via supersymmetry.   We view this as a ``CC miracle" in analogy to the WIMP miracle.    Secondly, we will show that the sector responsible for the crunching dynamics contributes to the  effective number of relativistic degrees of freedom,  $N_{\rm eff}$,  which is  expected to show up in future CMB experiments.   The absence of new physics at the LHC and the strong constraints on $N_{\rm eff}$~\cite{Aghanim:2018eyx} already introduce some tension with our scenario, suggesting a mild little hierarchy problem for the CC, analogous to the little hierarchy problem for the MSSM which was identified at the end of LEP and followed from  the direct searches and indirect precisions measurements at the time~\cite{Barbieri:2000gf}.

We present a concrete dynamical model that allows for the crunching of patches with large $\Lambda^4$. 
At  the core of the model is a spontaneously broken hidden conformal field theory (CFT) which does not contribute to the CC in the unbroken (high temperature) phase, while having a large   negative CC in its broken phase.   In the gravity dual, the CFT can be described via a stabilized Randall-Sundrum (RS) construction.
After inflation, the visible and hidden sectors are reheated, driving the CFT to its high temperature phase.   The crunching mechanism 
corresponds to  the subsequent supercooled phase transition of the CFT, which takes place at temperatures lower than the current temperature  of our observable universe.

The paper is organized as follows. In Sec.~\ref{sec:concept} we present the basic concept of our crunching mechanism and explain the three sectors needed together with their associated energy scales. In Sec.~\ref{sec:crunching} we introduce the hidden spontaneously broken CFT which will be used for the crunching sector, explain the two phases and their effective descriptions. We discuss the most important aspects of the crunching phase transition in Sec.~\ref{sec:PT} including both the O(3) and O(4) symmetric bubbles. For the O(3) symmetric bounce action we can use different approximations, while for the O(4) symmetric one relevant at low temperatures we rely on dimensional estimates. We show in Sec.~\ref{sec:CCmax} how the various cosmological constraints
limit the maximal CC that can be crunched away. The phenomenological consequences of our mechanism are discussed in Sec.~\ref{sec:pheno}, while we comment on the basic properties of the scanning and inflationary sectors in Sec.~\ref{sec:scanning}. Our conclusions as well as future prospects can be found in Sec.~\ref{sec:outlook}. The appendices contain a summary of the Goldberger-Wise  stabilization mechanism of the RS construction (App.~\ref{app:GW}), the discussion of the effects of an additional bulk gauge group on the dilaton potential (App.~\ref{app:QCD}), and more details on the various estimates of the bubble nucleation rate (App.~\ref{app:fullcalc}). 

\section{Basic Concept} 
\label{sec:concept}

In this section we describe the basic idea to address the CC problem.   We circumvent Weinberg's no go theorem by taking a similar approach to the anthropic solution~\cite{Weinberg:1987dv,Weinberg:1988cp}, namely we  assume that a large number of Hubble patches, each with a different value for the cosmological constant, are generated during inflation.
As opposed to the original idea, however, additional dynamics act to remove any patch with a large CC by causing it to crunch early during the cosmological evolution.  Consequently, only patches with small CC survive until today.   As a corollary, inflation need not be eternal or even last very long (see Sec.~\ref{sec:scanning}).  We shall further see that several observational consequences follow.  
During an initial inflationary  period,  different Hubble patches admit distinct values of the CC due to the presence of a scanning dynamics.  Once inflation ends, the universe with  all its patches  is reheated.
  In a given patch, the presence of a vacuum energy density of size $\Lambda^4$, triggers a phase transition at time of order $H_\Lambda^{-1} \sim \left(\Lambda^2/\sqrt{3} \mpl\right)^{-1}$, 
changing its value to be large and negative and  subsequently driving the patch to crunch.   The smaller the value of the CC, the longer the patch survives before crunching.   All patches crunch at finite time.

\begin{figure}[t]
\begin{center}
\includegraphics[width=7.1cm]{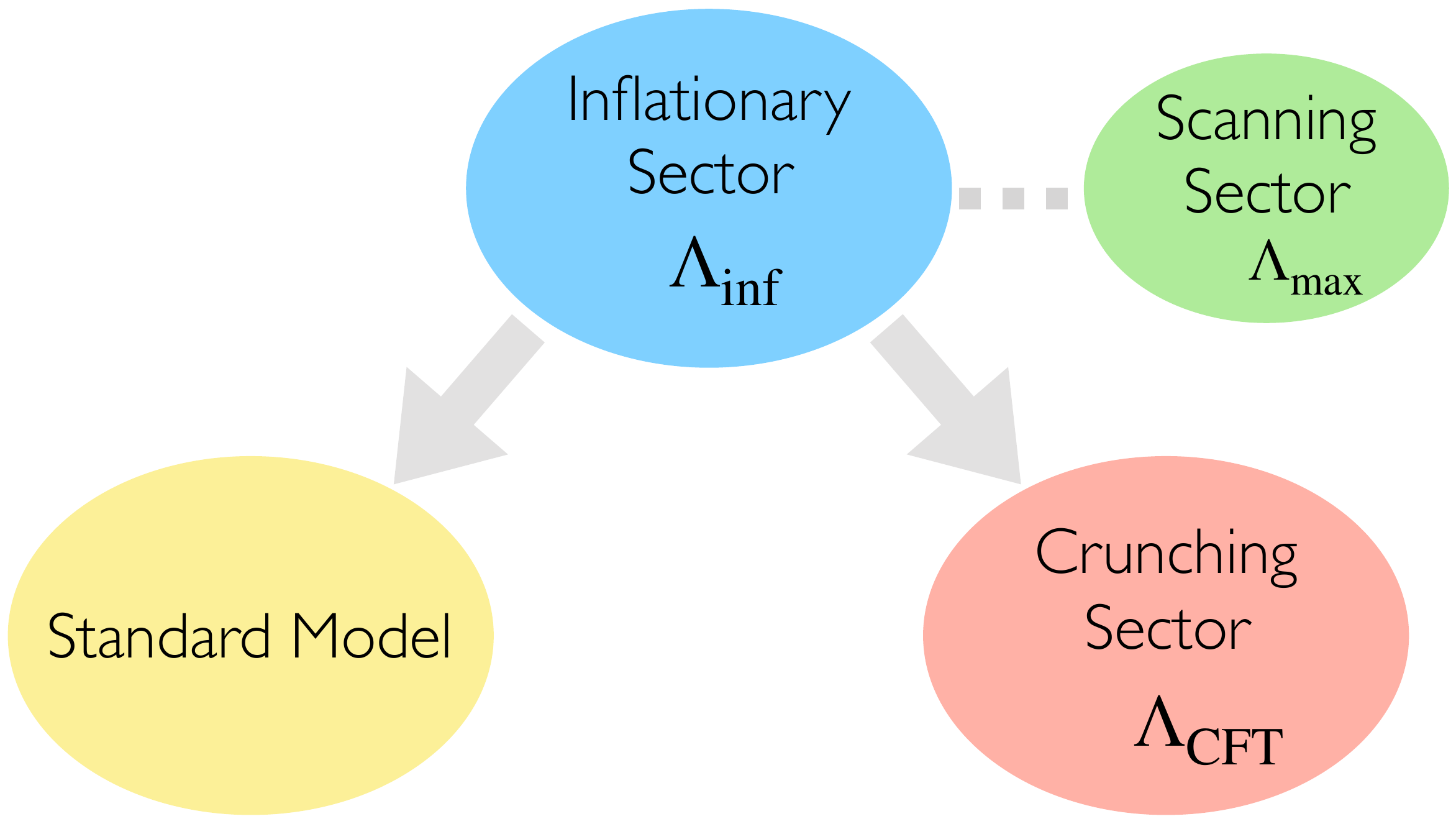}
\hspace{.5cm}
\includegraphics[width=7.2cm]{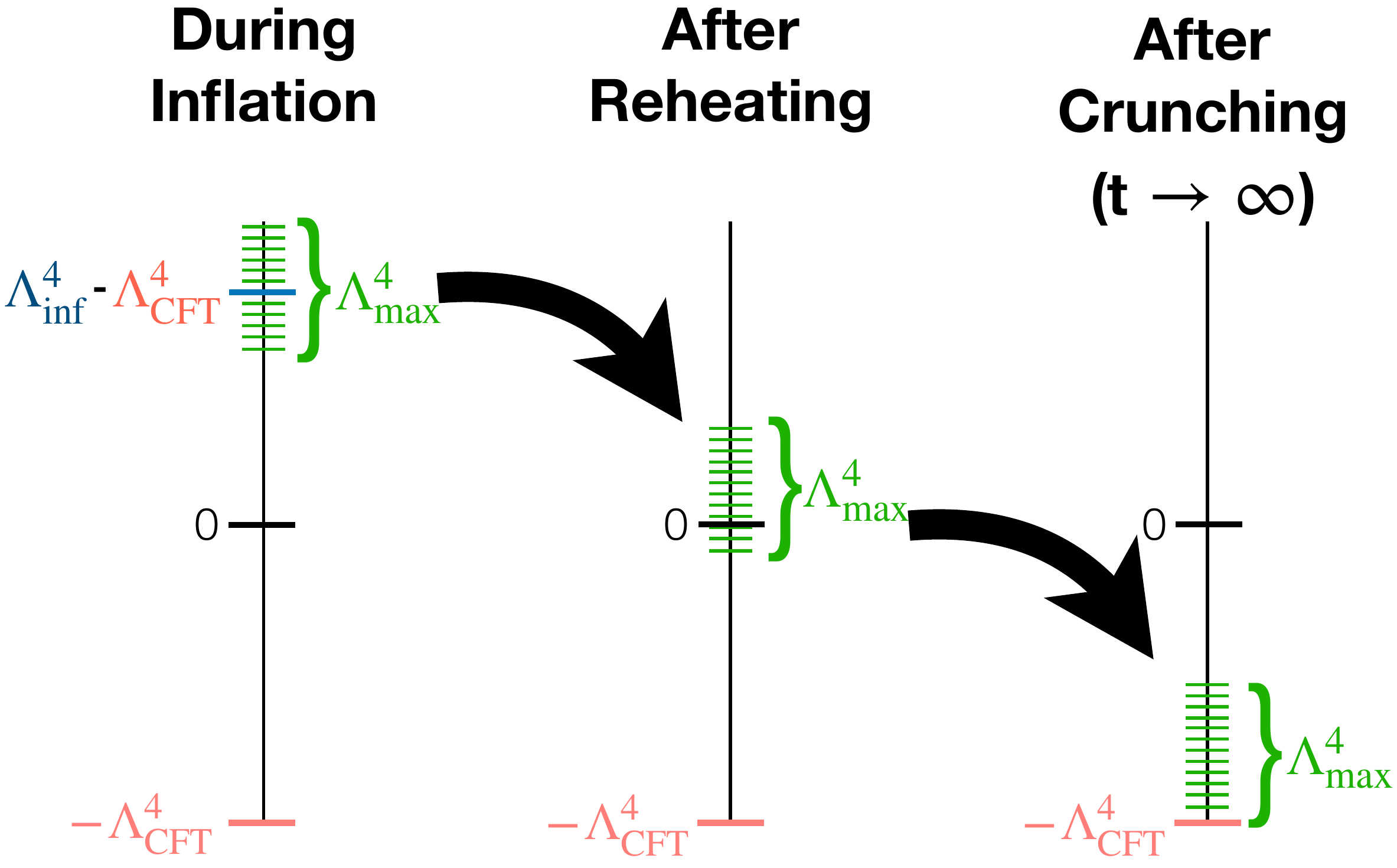}
\caption{\label{fig:scales} 
{\bf Left:} Our model assumes three different sectors in addition to the SM.   The inflationary sector is responsible for driving inflation at a scale $\Lambda_{\rm inf}$.  Inflation need not be eternal or have a large number of e-folds (see Sec.~\ref{sec:scanning} for the discussion on a lower bound).   The scanning sector may be independent of all other sectors and is responsible for producing a variety  of  Hubble patches with varying CC values.  
 The crunching sector, which takes the focus in this paper, is responsible for the dynamics that drives each patch with a large CC, $\Lambda^4 \leq \Lambda_{\rm max}^4$  to crunch within a time of order $H_\Lambda^{-1} \sim (\Lambda^2/\sqrt{3}\mpl)^{-1}$ from the end of the inflationary period.  In the scenario discussed below, this sector is a CFT, spontaneously broken at low temperature with a vacuum energy $-\Lambda_{\rm CFT}^4$.  {\bf Right:} The three scales are assumed to be (possibly very mildly) hierarchical: $\Lambda_{\rm max} < \Lambda_{\rm CFT} < \Lambda_{\rm inf}$.  During inflation the crunching sector sits in the true vacuum and the inflaton dominates the energy density.  A  variety of Hubble patches with energy densities lying around the scale $\Lambda_{\rm inf}^4 - \Lambda_{\rm CFT}^4$ and with a range of order $\Lambda_{\rm max}^4$, are populated during this epoch.   After reheating, the crunching sector sits in the false vacuum and does not contribute to the CC. The patches' energy densities are then distributed (possibly non-symmetrically) around zero with the same range of order $\Lambda_{\rm max}^4$.   Those with large negative cosmological constant will crunch independently of the crunching sector while those with positive (or small negative)  crunch only once the CFT phase transition is completed, thereby reducing their energy density by $\Lambda_{\rm CFT}^4$ and driving it to negative values.  Eventually, at late times, all patches go through the phase transition and crunch.
}
\end{center}
\end{figure}

Three somewhat independent modules of the theory, depicted in Fig~\ref{fig:scales}, are needed to achieve this scenario:
\begin{enumerate}
\item {\bf An inflationary sector} with scale $\Lambda_{\rm inf}$.  This sector drives a primary phase of inflation for {\it a finite time}, before reheating our universe.  During inflation, the inflaton's energy density is of order $\Lambda_{\rm inf}$.
\item {\bf A scanning sector} with scale $\Lambda_{\rm max}$.  Responsible for varying the CC during the initial inflationary period, thereby generating a multitude of domains with a landscape of contributions to the CC with a range $\Lambda_{\rm max}$.
\item {\bf A crunching sector} with a scale $\Lambda_{\rm CFT}$.  The dynamics in this sector reacts to the presence of a large CC and acts to lower its value, thereby driving the relevant patch to crunch.   In this paper this sector is a CFT spontaneously broken at low temperatures and with a vacuum energy density of order $-\Lambda_{\rm CFT}^4$.
\end{enumerate}
The sectors are not necessarily disconnected.   For example, in the model we describe below the crunching sector is assumed to couple to the inflaton which reheats it at the end of inflation.     In order to trigger the crunching of  every large CC region we assume $\Lambda_{\rm max} < \Lambda_{\rm CFT}$.   Conversely,  to allow for the primary phase of inflation (which occurs at zero temperature when the CFT is spontaneously broken), we must have $\Lambda_{\rm CFT} < \Lambda_{\rm inf}$.    Thus, all in all we assume the (possibly mild hierarchy), $\Lambda_{\rm max} < \Lambda_{\rm CFT} < \Lambda_{\rm inf}$ (see the right of Fig.~\ref{fig:scales}).   Note further, that the contribution of the inflationary sector to the energy density after inflation, is assumed to be smaller than $\Lambda_{\rm max}^4$, so that the observed value of the CC lies well within the landscape of vacua.

The inflationary and scanning sectors have been thoroughly studied in the literature (see e.g. Ref.~\cite{Starobinsky:1980te,Guth:1980zm,Linde:1981mu,Martin:2013tda, Brown:1987dd, Brown:1988kg, Arvanitaki:2009fg, Bousso:2000xa,Bachlechner:2017zpb}) and in this paper we are not adding anything new to that discussion.  In Sec.~\ref{sec:scanning} we shortly discuss  these sectors and derive a lower bound on the number of e-folds produced during the initial phase of inflation.

Several  approaches to the dynamics that drive the crunching regions may be taken.   Quite broadly such dynamics arise from either a CC-dependent field potential, or a CC-dominated cosmology which can turn the CC negative.
In this paper we take the latter approach.  At the end of inflation the crunching sector is assumed to be reheated and thereby driven into a long-lived false vacuum.  We then utilize the secondary phase of inflation triggered by the  CC in a given patch, setting off a phase transition.  Indeed, during inflation the temperature drops rapidly, going within a few Hubble times well below the critical temperature.   The decay of the (supercooled) false vacuum in the crunching sector decreases the CC to a negative value, driving to a short crunching period.   An exploration of a different scenario using the first approach is deferred for future work.

The crunching sector itself has non-trivial requirements which make it quite unique.
First, it needs to turn a large CC negative - so that the contribution to the CC from this dynamics should be large. However our patch of the Universe should not have crunched yet, hence the dynamics has to be such that the metauniversestable vacuum remains stable down to a very low temperature of order meV.  This means that the crunching sector in our patch is currently in a metastable state, and the difference in the vacuum energy between the metastable and stable minima is much larger than the characteristic temperature and energy scales in the universe when the transition occurs. This implies that the crunching sector must contain a supercooled phase in which the phase transition happens at a much lower temperature than the critical temperature.  

The outline for the history of the desired supercooled phase transition can be summarized in the following. The crunching sector is reheated above the critical temperature and remains in the false vacuum even below it. The transition occurs through bubble nucleation, with the nucleation temperature chosen to be below the current temperature of our observable universe. 
In areas with a large CC, the universe re-enters a secondary phase of inflation early on and the temperature of this sector drops exponentially with time until it reaches the nucleation temperature and triggers the transition. As a result, the CC turns negative within several Hubble times for every patch of the universe that re-enters inflation - and the longevity of each patch is inversely related to the value of its CC.

Our proposed dynamics solves all of the three dark-energy related questions discussed in the introduction. 
In the realization of the crunching dynamics we describe below, the coincidence problem (a.k.a.~the ``why now?" puzzle)  is solved by ensuring  our universe  lies close to a critical point, predicting our observable universe (and most others) to be on a verge of a catastrophic phase transition. 
Most importantly, our mechanism predicts new physics at the weak  scale, hinting at  a connection between the CC and the TeV scale.  At the same time it avoids  eternal inflation with its pitfalls, since all  patches in the universe are predicted to crunch thereby naturally circumventing the measure problem.  As we shall see, unlike the standard anthropic approach, our scenario is predictive and falsifiable.
We now move on to explain these statements in detail.

\section{The Crunching Sector}
\label{sec:crunching}

The essential new ingredient in our setup is the crunching sector that will effectively turn a large positive CC into a negative one. As discussed above, this sector is subject to two essential requirements: (i) an unstable  supercooled phase with a low nucleation  temperature,
and (ii) a true minimum with a large and negative energy density, $-\Lambda_{\rm CFT}^4$.  
 
 It is well-known that spontaneously broken conformal field theories (CFT's) feature exactly such a supercooled phase transition (at least for large N)~\cite{Witten:1998zw, Creminelli:2001th}.   In addition,  the CFT contribution to the CC from the unbroken phase vanishes, while in the broken phase it will be large and negative, providing the necessary jump in the CC to induce the crunch. The first order phase transition occurs via bubble nucleation~\cite{Callan:1977pt,Coleman:1977py,Linde:1981zj}. For a successful model, the tunneling probability from the metastable high-T vacuum to the true vacuum with broken scale invariance has to remain negligibly small down to $T \sim $meV, but allow for a transition not too far below it.   Thus at that temperature the tunneling probability should rapidly increase, thereby facilitating the phase transition with a large change in the vacuum energy. We note that an RS-type gravity dual of a large-N non-supersymmetric CFT~\cite{Randall:1999ee} satisfies (almost) all requirements and in particular, it features a first order  phase transition that is strongly supercooled.  As we shall see below, the shortcoming of a minimal RS model is that the bounce action (which controls  the tunneling probability), is only mildly temperature-dependent, implying a slow change in the probability.   To achieve a strong dependence we   introduce an explicit breaking of the CFT through a hidden QCD-like gauge theory in the bulk (along the lines of~\cite{vonHarling:2017yew, Baratella:2018pxi}), allowing for a phase transition soon below the meV scale.

\subsection{RS  at Zero Temperature}
\label{sec:RS0T}

For the concrete model of the  supercooled phase transition in cosmology we  use a minimal RS model with Goldberger-Wise (GW) stabilization~\cite{Goldberger:1999uk}, which  is dual to a spontaneously broken conformal field theory (CFT). In this model we have a slice of AdS$_5$ space described by the metric 
\begin{equation}
ds^2 = \frac{1}{k^2z^2} \left(dt^2-dz^2-
\sum_i dx_i^2\right)\,,
\label{eq:AdS}
\end{equation}
where $k$ is the AdS curvature. The AdS space is truncated at $z=1/k$ by the UV brane while at zero temperature there is an IR brane at $z=z_{\rm IR} \gg 1/k$. The position of this IR brane (in the absence of stabilization) is arbitrary, corresponding to the dilaton, $\chi \equiv 1/z_{\rm IR}$, the pseudo-Goldstone boson (pGB) of  broken  scale invariance. We stress that with our choice of parametrization, the dilaton is non-canonical,
\begin{equation}
\label{eq:dilatonkinetic}
{\mathcal L} \supset -\frac{3 (N^2-1)}{4 \pi^2} (\partial_\mu \chi)^2\,,
\end{equation}
where $N$ is the number of colors in the dual (unbroken) CFT, which is related  to the RS parameters via, $N^2-1 = 16\pi^2 (M_*/k)^3 = 4c/\pi$ with $c$ its central charge and $M_*$ the 5D Planck scale.

The GW stabilization mechanism involves adding a massive bulk scalar field, $\phi$, with boundary potentials on the two branes. These  potentials together with the bulk mass,  forces a non-trivial 5D profile for the scalar, which in turn  results in the stabilization of the radion. For more details see App.~\ref{app:GW} and~\cite{Goldberger:1999uk, Csaki:2004ay}. We analyze the theory assuming that the dilaton is light compared to  all other composite degrees of freedom, and restrict ourselves to the effective potential of a stabilized dilaton (after integrating out the bulk scalar and graviton modes) of the form,
  \begin{equation}
 V_{\rm eff}(\chi)= -\lambda \chi^4 + \frac{\lambda_1}{k^{\epsilon_1}} \chi^{4+ \epsilon_1} - \frac{\lambda_2}{k^{-\epsilon_2} }\chi^{4-\epsilon_2} \label{eq:DilatonPotential}\, .
 \end{equation}
Here we have chosen the arbitrary normalization scale to be $k$ for simplicity.
 The $\chi^4$ term is expected to appear in the scale invariant theory, while the terms with powers $4\pm\epsilon_{1,2}$ correspond to explicit breaking of scale invariance.   For the case of the GW stabilization, the $\lambda$ coefficients are smaller than one but not necessarily hierarchically so, while  $\epsilon_i\ll1$, thereby generating the  UV-IR hierarchy (also in that case $\epsilon_2 \simeq -2\epsilon_1$).   As we show in Sec.~\ref{sec:PT},  a supercooled phase transition, (from a hot CFT phase to the spontaneously broken one),  with a  bounce action that quickly drops around some temperature, $T_*$, requires, for example, $\epsilon_2$ to be ${\cal O}(1)$ and positive.  Demanding further that  $T_* \ll k$ implies that $\lambda_2\ll 1$, (see Eqs.~\eqref{eq:chistar} and \eqref{eq:Tstar}).
  With this choice, the third term in Eq.~\eqref{eq:DilatonPotential} is negligible except for very small $\chi$ and in particular, it does not influence the location of the minimum which is given by 
  \begin{equation}
  \label{eq:GWminimum}
 \chi_{\rm min} \simeq  k \left(\frac{1}{(1+\epsilon_1/4)}\frac{\lambda}{\lambda_1} \right)^{1/\epsilon_{1}}.
 \end{equation}
The role of the third term in Eq.~\eqref{eq:DilatonPotential} is nonetheless crucial  due to its domination at low values of $\chi$.  For $\chi$ below the critical value, $\chi_*$,  
\begin{equation}
\label{eq:chistar}
\chi_* \equiv \left(\frac{2\pi^2(4-\epsilon_2)(3-\epsilon_2) \lambda_2}{3(N^2-1)}\right)^{1/\epsilon_2} k \ll \chi_{\rm min}\,,
\end{equation}
the mass of the dilaton becomes larger than the IR scale and the effective theory breaks down as the dilaton is no longer lighter than the other KK-states.   Around $\chi_*$ the third term therefore dominates the potential and affects crucially  the nucleation rate and its dependence on the  temperature. 
We describe this effect in the next section.  A more detailed explanation of the setup for this model is presented in appendices~\ref{app:GW} and~\ref{app:QCD}.

 As was  shown in~\cite{vonHarling:2017yew,Baratella:2018pxi} and reviewed in~App.~\ref{app:QCD}, the third term of Eq.~\eqref{eq:DilatonPotential} with the properties discussed above (i.e.~$\epsilon_2$ of order unity and  $\lambda_2\ll1$), can originate from an additional gauge group in the bulk of the AdS.  In particular, the RG flow of the gauge group, and correspondingly the confining scale, $\Lambda^\prime$, are influenced by the CFT degrees of freedom.  When the confinement occurs below the  CFT breaking scale, $\Lambda^\prime$ becomes dependent on the dilaton (which parametrizes the CFT breaking), $\Lambda^\prime=\Lambda^\prime(\chi)$.  As a result, the confinement  energy density, which is an explicit breaking of the CFT, introduces the necessary effective potential for the dilaton.

\subsection{RS at Finite Temperature}

So far, we have discussed the GW stabilization which describes the true vacuum of the theory.   This corresponds to the stable phase of the system at low temperatures.  
A second phase of the theory is apparent at finite temperature.  This phase  corresponds to the symmetric CFT phase and is described on the 5D side  by the Euclidean AdS-Schwarzschild (AdS-S) solution given by the metric~\cite{Creminelli:2001th}
 \begin{equation}
ds^2 = \left( \frac{1}{k^2z^2} - \frac{z^2}{k^2 z_H^4}\right) dt^2 +\left(\frac{1}{k^2z^2} - \frac{z^2 }{k^2 z_H^4}\right)^{-1} \frac{ dz^2}{k^4z^4}  + \frac{1}{k^2z^2} \sum_i dx_i^2 \,,
\label{eq:AdSS}
\end{equation}
where $z_H$ corresponds to  the location of a black brane (BB) horizon. For $z_H\to \infty$  the AdS metric, Eq.~\eqref{eq:AdS}, is recovered. A corresponding Hawking temperature of the BB can be defined and is given by
\begin{equation}
T_H = \frac{1}{\pi z_H}\,.
\end{equation}
At equilibrium, $T_h$ is set by the temperature of the system, which also defines the periodicity of the Euclidean time direction, $T_H =  T$.
For other choices of $T_h$, the system is out of equilibrium (or equivalently, away from the minimum of the potential), and a conical singularity at the horizon appears.

The transition between the two space-times is known as the Hawking-Page phase transition~\cite{Hawking:1982dh}. 
 The properties of this phase transition are easier to understand from the CFT side. At low temperatures the theory describes a spontaneously broken conformal symmetry, while at high temperatures, the symmetry is restored.  The critical temperature of the  phase transition is defined by the equality of the free energies of the two phases. The free energy of the broken, low temperature phase, is dominated by potential energy,
\begin{equation}
F_{\rm broken}(\chi,T) \simeq V(\chi) \,, 
\end{equation}
and in particular the minimum has $\Lambda_{\rm CFT}^4 \simeq -F_{\rm broken}(\chi_{\rm min}) \simeq \frac{\epsilon_1 \lambda}{4}  \chi^4_{\rm min}$.   Here $\Lambda_{\rm CFT}$ is the relevant  energy scale of the crunching sector discussed in Sec.~\ref{sec:concept}.  The hot conformal phase is simply black-body radiation with free energy
\begin{equation}
\label{eq:Fconf}
F_{\rm conformal} = - \frac{\pi^2}{8} N^2 T^4\,.
\end{equation}
 The critical temperature is therefore found to be
\begin{equation}
T_c = \chi_{\rm min} \left(\frac{2\epsilon_1\lambda}{\pi^2N^2}\right)^{1/4}
\end{equation}
As we discuss in the  next section, the phase transition occurs at temperatures much below the  critical one.

To describe the phase transition, one must identify the relevant degrees of freedom in each of the phases.   In the broken phase, and sufficiently far from the origin, $\chi \gsim  T\max[1,  k/M_*]$, no KK states are dynamical and quantum gravity  corrections are subdominant and therefore the system is adequately described by the light dilaton (after canonically  normalizing) with $V(\chi)$ given by Eq.~\eqref{eq:DilatonPotential}.   For $\chi \lesssim T\max[1,  k/M_*]$ however, the effective potential will get large T-dependent non-calculable corrections and additional degrees of freedom are expected to be needed for the description of the theory.   Meanwhile, in the absence of scales in the hot CFT, the effective potential in the unbroken phase is unknown, and again many degrees of freedom are expected to be dynamical.  On the gravity side one can identify the inverse location of the BB horizon or equivalently its Hawking temperature, $T_h$, as a degree of freedom which controls the free energy of the system at the minimum (as well as the height of the barrier).  We stress, however, that $T_h$  is non-canonical and may be accompanied by many other degrees of freedom.   Nonetheless we use $T_h$ as a convenient illustrative parametrization of the physics of the hot CFT phase.

\subsection{The Potential at Small \texorpdfstring{$\chi$}{chi}}

\begin{figure}
\begin{center}
\includegraphics[width=10cm]{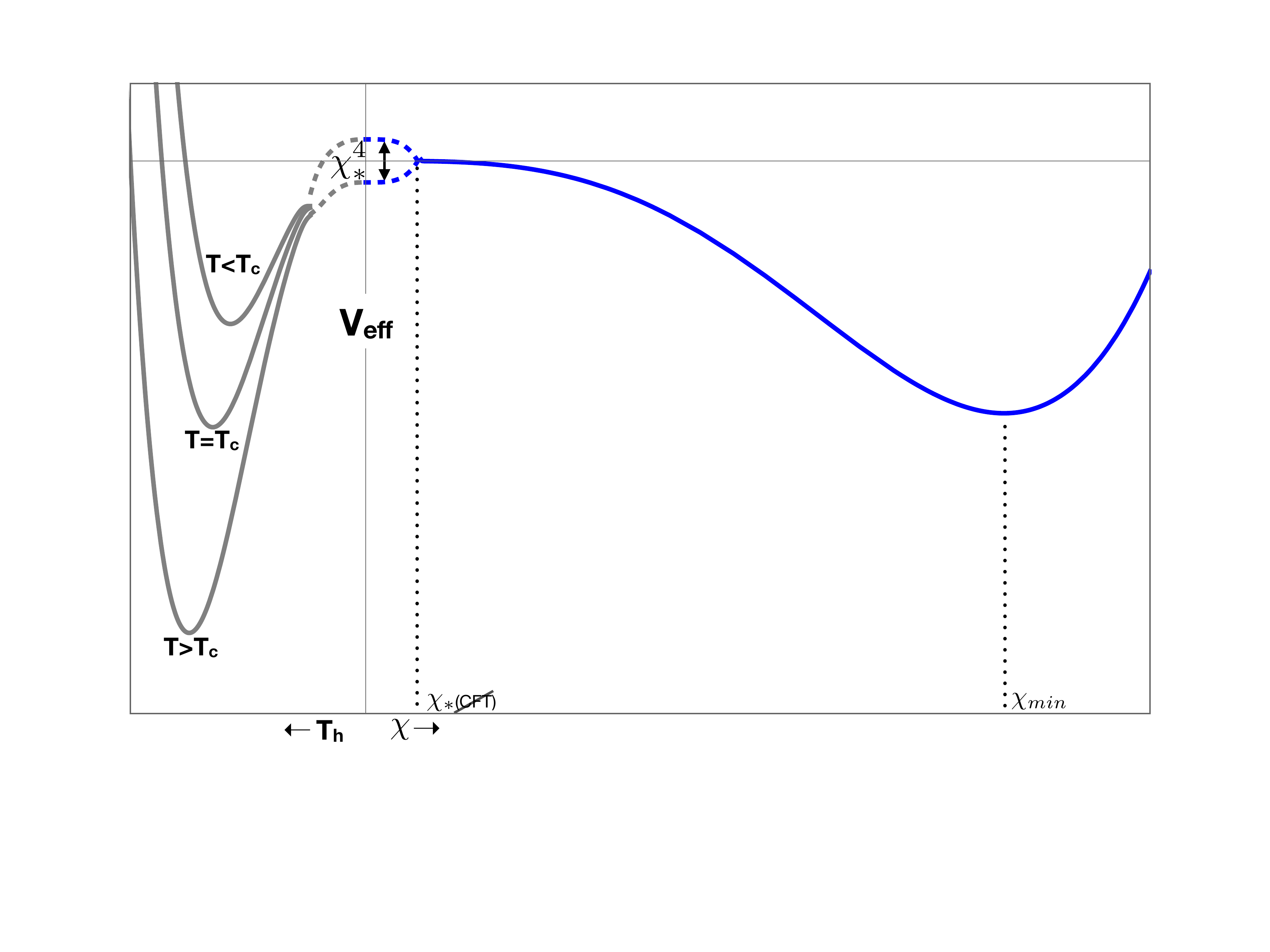}
\caption{A 1D illustration of the potential describing the two phases of the crunching CFT sector. The potential is obtained by gluing together the zero temperature dilaton potential, $V(\chi)$
(on the right of the vertical line at $\chi = T_h = 0$) and the thermal potential of the black brane as a function of its Hawking temperate, $T_h$ (on the left). The gluing is done when both the IR brane and the black brane horizons are taken to infinity. At large $\chi$, the Goldberger-Wise potential is dominating.  At    $\chi \simeq \chi_*$ and below,  the explicit breaking of the conformal symmetry becomes sizeable and deforms the potential, introducing a barrier with a typical scale $\chi_*$.    In our scenario, the explicit breaking arises from an additional confining gauge group in the bulk of the AdS.
\label{fig:DilatonPotential}}
\end{center}
\end{figure}

\begin{figure}[t]
\begin{center}
\includegraphics[width=14cm]{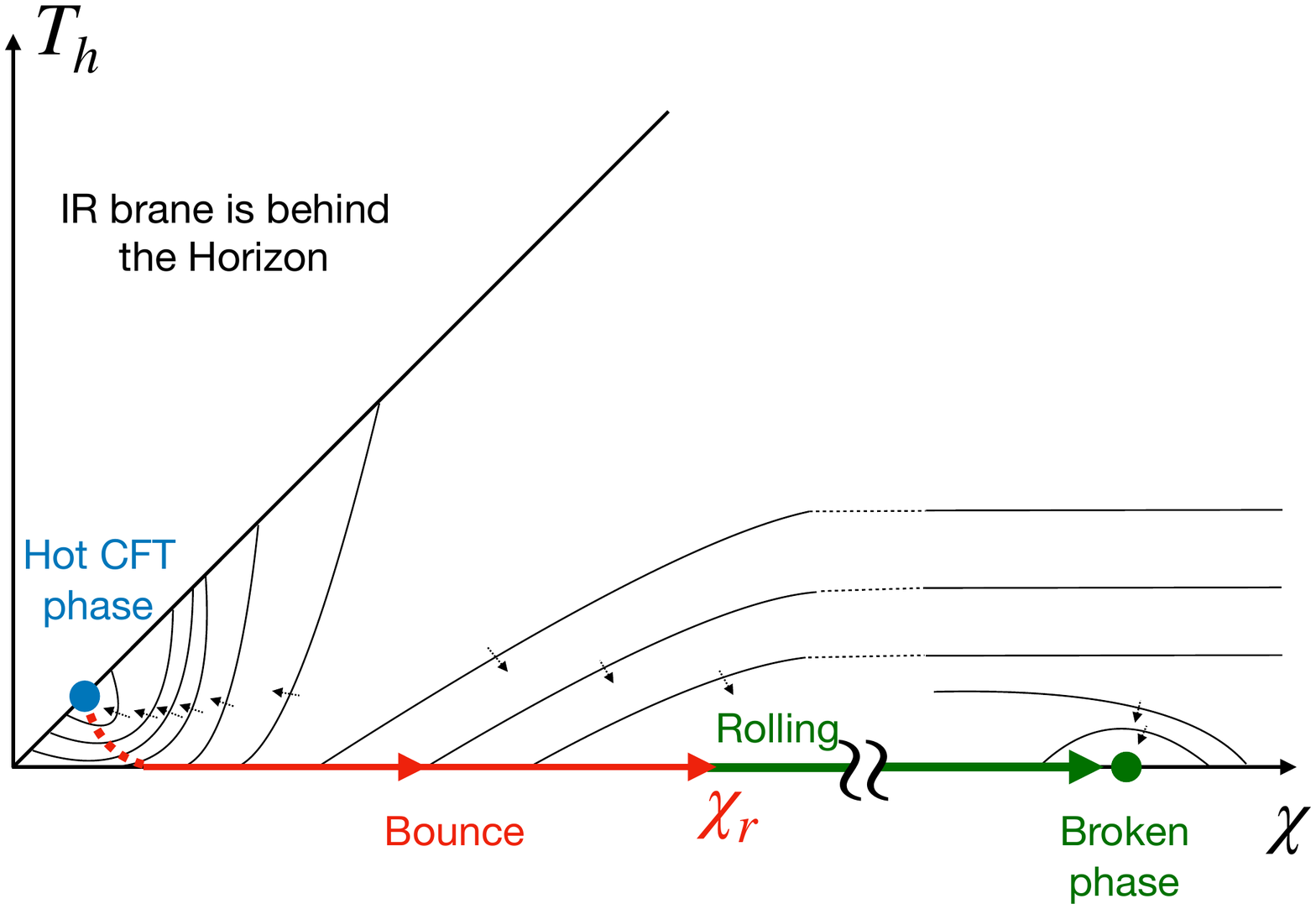}
\caption{\label{fig:Roads}A 2D illustration of the effective potential describing both phases of the CFT: the spontaneously broken phase parametrized by the dilaton, $\chi$, and the hot conformally-restored phase parametrized by the horizon Hawking temperature of the 5D black brane dual.  $\chi=0$ corresponds in the RS picture (i.e.~the spontaneously broken phase) to the IR brane at infinity.  Conversely, the AdS-S geometry, dual to the hot CFT phase, is truncated at the horizon and the IR brane can be thought of sitting there and contributing no energy due to the vanishing warp factor.   We stress that $T_h$ is not a  canonical degree of freedom nor is it parametrically lighter than any other excitation in that phase.  Consequently the (multi-dimensional) potential is unknown in this regime.  Furthermore, in the spontaneously broken phase, the effective theory breaks down for $\chi<T\max[1,  k/M_*]$, where additional KK modes are expected to enter and/or non-calculable quantum gravity corrections must be taken into account.  
Below the critical temperature, the spontaneously broken phase lies parametrically farther from the origin in field space compared with the hot CFT minimum, since the relevant scale in the latter phase is the temperature $T$.
This fact plays a role in the calculation of the bounce action, which for sufficiently large temperatures is dominated by the semi-classical contribution  from the broken phase.
The red line shows the path of an effective euclidean particle  tunneling from the false vacuum to the true one.  The solid part describes the dominant and calculable contribution  while the dotted part illustrates the non-calculable but  subdominant part of the action for $\chi_r >T\max[1,  k/M_*]$, where $\chi_r$ is the release point  of the bounce.  The green line illustrates the Lorentzian-time  path of the field, as it rolls down to the minimum of the broken phase after tunneling.}
\end{center}
\end{figure}

Before moving on to the dynamics of the phase transition we would like to discuss the shape of the dilaton potential, and in particular its behavior for small values of $\chi$. 
A one- and  a two-dimensional illustration of the potential as parametrized by $\chi$ and $T_h$ are presented in Figs.~\ref{fig:DilatonPotential} and~\ref{fig:Roads}.
 The 1D  illustration of Fig.~\ref{fig:DilatonPotential}  was obtained by gluing the dilaton potential to a finite temperature potential for the BB horizon, where the gluing is done at the point where both  are taken to infinity. This is the approach taken in \cite{Creminelli:2001th}, where the authors use the value of the free energy for an unbroken CFT at zero temperature as their reference, setting it to zero. In \cite{Creminelli:2001th}, the 1D parameterization correctly reproduces the size of the barrier and shows that the potential is much steeper on the BB side. 
 A slightly more informative description is given by a two dimensional plot in Fig.~\ref{fig:Roads}, where we illustrate the potential as a function of the IR brane and the horizon location.

Our analysis here differs from \cite{Creminelli:2001th} in one important detail.  We have seen that at a value of the dilaton $\chi=\chi_*$ the additional term $\propto \chi^{4-\epsilon_2}$ in Eq.~\eqref{eq:DilatonPotential} will become dominant. In the CFT language this implies that the effect of an explicit breaking term is becoming ${\cal O}(1)$ and one expects the RG flow to be driven to a different fixed point. In the 5D AdS language this would be described by a domain wall at a position around $\chi=\chi_*$  that will separate the AdS space from whatever space arises as a result of the backreaction (possibly  another AdS space asymptotically). While we do not know much about the details of the behavior of the theory below $\chi_*$,  we do expect  that there will be a term in the potential set by $\chi_*$, which can be understood either as a threshold correction in the CFT language or the contribution of the domain wall sitting at $\chi_*$ in the 5D language. An illustration of this effect is given in Fig.~\ref{fig:DilatonPotential}, taking either a negative or positive tension for the domain wall.
This effect generates a barrier of order $\chi_*^4$ in the dilaton potential, implying that the non-calculable regime $\chi < \chi_*$ will be dominated by the energy scale $\chi_*$, and consequently dimensional analysis suggest that  the bubble nucleation probability is also  set by the same scale.

\section{A Crunching Phase Transition\label{sec:PT}}

\subsection{Preliminaries}

The essence of our crunching mechanism is the Hawking-Page phase transition~\cite{Hawking:1982dh} corresponding to the decay of the metastable high-temperature CFT (described by the black brane of the AdS-S metric~\cite{Creminelli:2001th}) to the spontaneously broken minimum (described by the RS with the GW solution~\cite{Goldberger:1999uk}).  Usually two contributions to the bubble nucleation rate play a role:  an $O(4)$-invariant action~\cite{Coleman:1977py}, $S_4$, and an $O(3)$-invariant finite-temperature contribution, $S_3(T)/T$~\cite{Linde:1981zj}.   Here we set the stage to the estimation of both.

\subsubsection*{The $O(3)$ Action}
To calculate $S_3(T)$, one needs to solve the 4-dimensional Euclidean classical EOM, where it is assumed that the solution is independent of the compactified Euclidean time, and spherically symmetric in the other three. For a canonically normalized field $\phi$, the EOMs in terms of the 3D radius is,
\begin{equation}
\label{eq:S3EOM}
\phi^{\prime \prime}+\frac{2}{r} \phi^{\prime} -V^{\prime}(\phi, T)=0\,,
\end{equation}
where $V(\phi, T)$ is the potential describing the system.
This equation can be understood as describing the classical motion of a point particle in an inverted potential $-V$ and a friction term, with $r$ playing the role of time.
The corresponding bounce action is then given by,
\begin{equation}
\label{eq:S3}
S_{3}(T)=4\pi\int^{\infty}_0dr\cdot r^2\left[\frac{\bar\phi'^{2}}{2}+\overline V(\bar\phi, T)\right]\,,
\end{equation}
where $\overline V(\phi, T)$ is the potential shifted to vanish in the false vacuum and  $\bar\phi$ is the solution to the equation of motion, Eq.~\eqref{eq:S3EOM}, with the boundary conditions $\bar\phi'(r=0)=0$ and $\bar\phi(r\rightarrow\infty) = \phi_{\rm false}(T)$. The value at the origin $\bar\phi(r=0)\equiv \phi_r(T)$ is referred to as the release value of the field $\phi$, which sits on the true vacuum side of the barrier, while $\phi_{\rm false}(T)$ is the value of the field at the false  vacuum, corresponding to the  hot CFT phase in our case.

As we have seen, our system exhibits two different phases with very distinct sets of degrees of freedom.  What should one use for the corresponding field, $\phi$? 
In~\cite{Creminelli:2001th} the bounce was estimated by gluing the potential in the AdS-S solution with the horizon at infinity, to the GW potential in the RS solution with the IR brane at infinity. The instanton trajectory would correspond to the motion of the AdS-S horizon, parametrized by $T_h$, from the false vacuum to infinity, followed by the motion of the IR brane (described by its location $z_{\rm IR} = 1/\chi$),  from infinity ($\chi = 0$)  to the release point $\chi_r$.  In Fig.~\ref{fig:Roads} this transition would be seen with the solid red line reaching all the way to the origin, followed by a path along the  $T_h=\chi$ line  to the minimum associated with the hot CFT phase.
It is however unlikely to be the precise path describing the minimal bounce action since  the dilaton effective potential breaks down for $\chi < T\max[1,  k/M_*]$ while the CFT potential is unknown and likely multi-dimensional. The bounce action therefore consists of a calculable and a non-calculable contribution,
\begin{equation}
\label{eq:S3contributions}
S_3(T) = S_3^{\rm calc}(T) + S_3^{\rm non-calc}(T)\,.
\end{equation}
The solid red line in Fig.~\ref{fig:Roads} illustrates the calculable part of the bounce action while the dotted red line shows the non-calculable part.  

Since the  minimum of the broken phase lies far from the origin, $\chi_{\rm min} \gg T\max[1,  k/M_*]$ [see Eq.~\eqref{eq:GWminimum}], while the field space of the unbroken phase spans a distance of order $T$, one expects the calculable part of $S_3(T)$ to dominate over the non-calculable part.    Below we show that as the temperature drops, the release point, $\chi_r(T)$, associated with the boundary condition at the center of the bubble, moves towards the origin.  Once $\chi_r(T) \leq \chi_*$, the calculable part of the action vanishes.  
As we argue below and in App.~\ref{app:uncertainties}, the absence of a small parameter suggests that the (now dominating) non-calculable part of the action for such low temperatures is of order $N^2$, $S_3^{\rm non-calc}(T)/T = {\cal O}(N^2)$\footnote{Strictly speaking,  this N-dependence ignores non-perturbative corrections which may be large.}. A similar statement is true for the $S_4$  action discussed below.  We will use this fact when estimating the largest possible CC that can be crunched away using the dynamics discussed here.

At sufficiently high temperature, 
the bounce action discussed above governs the rate of thermal bubble nucleation per unit volume which is given by~\cite{Linde:1981zj}
\begin{equation}
\label{eq:decayrate}
\frac{\Gamma_3(T)}{V}=\Gamma_0^{(3)} \cdot T^{4}\left(\frac{S_{3}(T)}{2 \pi T}\right)^{3 / 2}  e^{-S_{3}(T) / T}  \qquad  \textrm{for\,\,\, } T\gg T_*\,,
\end{equation}
where $\Gamma_0^{(3)}$ is an ${\cal O}(10^{-3}-10^3)$ pre-factor, as discussed in App.~\ref{app:uncertainties}.  The $T^4$ prefactor is easily understood: for high $T$ the theory is approximately conformal and the only explicit scale in the theory is $T$.  
As the temperature drops and the release point for the bounce solution, $\chi_r$, reaches $\chi_*$, conformality is explicitly broken and 
the characteristic scale is set by $\chi_*$. At those low temperatures we expect the $O(4)$-invariant bubble to become dominant, but since we have seen that the bounce action is no longer large nor calculable, the prefactor is expected to be on dimensional grounds be of order $\chi_*^4$,  $\Gamma(T\lesssim T_*)/V \propto \chi_*^4$.

 \subsubsection*{The $O(4)$ Action\label{sec:S4}}

So far, we have discussed the finite-temperature $O(3)$-symmetric contribution to the bounce action.   We now discuss the properties of the $O(4)$-symmetric solution.   To understand its importance, let us first note that the $S_4$ bounce action becomes dominant only when two conditions are met~\cite{Linde:1981zj}, (i) $R_{\rm bubble} T\leq 1$ with $R_{\rm bubble}$ the size of the  bubble, and (ii) $S_4\leq S_3(T)/T$. Indeed, at some low temperature $T$, the bubble radius plateaus due to the finite $\chi_*$-scale barrier, and as we continue to lower the temperature, $R_{\rm bubble} T\rightarrow 0$.  Additionally, as the temperature is sufficiently lowered, $S_3$ plateaus as well, while $T$ continues to decrease, thus for some low temperature we will clearly have $S_3(T)/T>S_4$. Therefore, one  finds that the O(4)-invariant solution will dominate the bubble nucleation rate.

As in the $O(3)$ case, the $O(4)$-symmetric action is given by
\begin{equation}
S_4=2\pi^2\int r^3 dr\left[\frac{\overline\phi^{\prime2}}{2} + \overline V(\overline\phi)\right]\,,
\end{equation}
where as before, $\overline\phi$ denotes the solution to the Euclidean EOM (this time for a 4D invariant solution)
\begin{equation}
\label{eq:S4EOM}
 \phi^{\prime \prime}+\frac{3}{r} \phi^{\prime} -V^{\prime}(\phi, T)=0\,
\end{equation}
and  $\overline V$ denotes the potential shifted to zero in the false vacuum.    Once again, the field $\bar\phi$ will be taken to be the dilaton in the broken phase.  
Finally,  the corresponding rate can  be estimated using the characteristic scale, $\chi_*$, which is the only dimensionful parameter relevant for the $O(4)$ bounce solution~\cite{Callan:1977pt}
\begin{equation}
\label{eq:O4decayrate}
\frac{\Gamma_4(T) }{V}= \Gamma_0^{(4)}\chi_*^4 \left(\frac{S_4}{2\pi}\right)^2 e^{-S_{4}}\,.  
\end{equation}
As with $\Gamma_0^{(3)}$, here $\Gamma_0^{(4)}$ parametrizes our ignorance and following~\cite{Linde:1981zj} taken to be of order ${\cal O}(10^{-3}-10^3)$ and ${\cal O}(10^{-4}-10^4)$ respectively (see App.~\ref{app:uncertainties}).

\subsection{Estimates for the $O(3)$ and $O(4)$ Actions}
\label{sec:estimates}

In order to gain intuition  
and identify the interesting regions of parameter space, it is useful to derive  approximate expressions for the parametric dependence of $S_3(T)/T$ and $S_4$.   In this section we find a back-of-the-envelope estimate, while a more careful treatment of the $S_3(T)/T$ is given in App.~\ref{app:fullcalc} where we numerically calculate the bounce action.  

We begin by identifying a characteristic temperature scale, $T_*$ (which we mentioned already in Sec.~\ref{sec:RS0T}),  at which the phase transition takes place.   This scale can be defined as the temperature such that 
\begin{equation}
\label{eq:TstarDef}
\chi_r(T=T_*)=\chi_*\,.
\end{equation}
Recall that $\chi_*$ defines the value of $\chi$ for which 
the explicit  breaking of conformal symmetry is sizable and 
the dilaton is no longer the only relevant degree of freedom.  Thus at and below $T_*$ the bounce solution runs along $\chi<\chi_*$ and the action is fully non-calculable.  
In accordance with the discussion above (and derivation below), due to  the absence of small parameters,  at this value $S_3(T_*)/T_* \simeq {\cal O}(N^{2})$ and the decay rate is sizable.   
As we show momentarily, at around $T_*$, $S_4$ becomes dominant and bubble nucleation proceeds via the zero-temperature solution, with similar scaling of the bounce-action.    The above implies that at around $T_*$  the decay rate peaks and  is indeed sizable, and therefore  bubble nucleation can occur allowing for the phase transition to take place.  

Another important scale is the  size of the nucleated bubble, $R_{\rm bubble}$.   
The typical distance scale associated with the solution of Eq.~\eqref{eq:S3EOM} [for $\phi\rightarrow\sqrt{3(N^2-1)/(2\pi^2)}\chi$ and $V \rightarrow V_{\rm eff}(\chi)$], gives a good estimate for the size of the bubble\footnote{$T$-dependence will only enter through the boundary conditions.} and scales as 
\begin{equation}
\label{eq:Rbubble}
R_{\rm bubble} \sim \frac{\sqrt{3}}{2\pi} N [V_{\rm eff}''(\chi_r)]^{-1/2} = \frac{1}{\sqrt{2}} N  \chi_r^{-1}\left[N^2\left(\frac{\chi_*}{\chi_r}\right)^{\epsilon_2}+ 8\pi^2\lambda \right]^{-1/2}  \,,
\end{equation}
where $\chi_*$ is defined in Eq.~\eqref{eq:chistar}.
 From the above we see that the last term of Eq.~\eqref{eq:DilatonPotential} will start to dominate 
for $\chi_r\simeq (\lambda/N^2)^{-1/\epsilon_2}\ \chi_* > \chi_*$.  Below we use $R_{\rm bubble}$ to show that around $T_*$, the phase transition proceeds via nucleation of $O(4)$-symmetric bubbles.

\subsubsection*{$O(3)$ Estimate}

Next, let us  present a back-of-an-envelope estimation for $S_3(T)/T$.   $\chi_r$ can be estimated using energy considerations: as $\chi$ rolls down the inverted $V_{\rm eff}$ potential from $\chi_r$ to the origin, it gains kinetic energy.  This kinetic energy must then be used to climb up to the peak of the  inverted potential corresponding to the false vacuum of the hot CFT of order $V_{\rm CFT} \sim \frac{\pi^2}{8} N^2 T^4$.  Neglecting the friction term in Eq.~\eqref{eq:S3EOM}, and equating the two energies $V_{\rm eff} \sim V_{\rm CFT}$ one finds
 \begin{equation}
\label{eq:chirT}
\chi_r \sim  \chi_* \ \textrm{min}\left[\left(\frac{T}{T_*}\right)^{\frac{1}{1-\epsilon_2/4}},  \left(\frac{3N^2}{2\pi^2 (4-\epsilon_2)(3-\epsilon_2) \lambda}\right)^{1/4}\frac{T}{T_*}\right]\,, 
\end{equation}
where we use an estimate,
\begin{equation}
\label{eq:Tstar}
T_* \sim \left(\frac{12}{(4-\epsilon_2)(3-\epsilon_2)\pi^4} \right)^{1/4} \chi_* \,.
\end{equation}

With  the above, we can infer the contribution to the action from the kinetic term in Eq.~\eqref{eq:S3}, by taking $\phi'^2\sim (3N^2)/(2\pi^2) \chi_r^2/R_{\rm bubble}^2$, and $\int r^2 dr\sim R_{\rm bubble}^3$. Using integration by parts and the EOM, Eq.~\eqref{eq:S3EOM}, we can rewrite the action only using the kinetic term, $S_3=(4\pi/3) \int r^2 dr \phi'^2$. We therefore approximate the action as,
\begin{equation}
\label{eq:S3estimate}
\frac{S_3(T)}{T}\sim\ \textrm{min}\left[ \left(\frac{(4-\epsilon_2)(3-\epsilon_2)}{3} \right)^{1/4} N^{2}\left(\frac{T}{T_*}\right)^{\frac{3\epsilon_2/4}{1-\epsilon_2/4}} , \frac{N^{7/2}}{ 2}\left(\frac{1}{2\pi^2 \lambda}\right)^{3/4} \right]\,.
\end{equation}
We stress that this estimate is for the action integrated all the way to zero, hence it includes a contribution from the non-calculable regime which should not be trusted. This contribution is however $\mathcal{O}(1)$ of the total action when the action is large -- making the back-of-the-envelope calculation a reasonably accurate estimate to the numerical calculation  presented in App.~\ref{app:numericbounce}, which excludes the non-calculable region.   Conversely, at around $T_*$, when the non-calculable contribution dominates, the above can significantly overestimate the true action and should  not be trusted, especially for small $\epsilon_2$.   The lower-bound numerical calculation of App.~\ref{app:numericbounce} without  the non-calculable regime, is then expected to give a closer estimation of the true bounce action.

Two comments are in order.   First, we see that at high enough temperatures, when the bounce action is dominated by the $\lambda\chi^4$ term, it no longer depends on $T$, as expected in a conformal theory. 
Second, as the temperature drops towards $T_*$, the release point $\chi_r$  is driven to $\chi_*$ from above.  Hence at  $T \gg T_*$ our leading order estimation may indeed be trusted.

\subsubsection*{$O(4)$ Estimate}

Moving  to estimate $S_4$, recall that the effective action is non-calculable for $\chi< \chi_*$ and the corresponding back reaction of the scale-invariant breaking source on the AdS geometry is expected to be sizable.  As a consequence we can not directly calculate the O(4)-symmetric rate,  however   we can build on the simple dimensional analysis used for Eq.~\eqref{eq:O4decayrate} to  estimate it.

Using  integration by parts and the EOMs, we find that the dynamics of the bounce solution admits virialization, that is (for O(4) symmetric bubbles) $\int\overline V(\overline\phi)\sim-\int\overline\phi^{\prime2}/4$. Therefore, in a similar fashion to the $S_3(T)/T$ back-of-an-envelope result, we can estimate $S_4\sim  2\pi^2(3N^2)/(4\pi^2) R_{\rm bubble}^2  \chi_r^2/4$.  Plugging our estimate for $R_{\rm bubble}$ from Eq.~\eqref{eq:Rbubble}, we find
\begin{equation}
\label{eq:S4envelope}
S_4 =\frac{3c_0}{16}N^2\left(\frac{\chi_r}{\chi_*}\right)^{\epsilon_2}\lsim \frac{3c_0}{16}N^2\, ,
\end{equation}
where $c_0$ is a fudge factor parameterizing our ignorance.  
Since in this regime the theory has no small parameters, the prefactor $c_0$ in the above  is expected to be order one.  In Fig.~\ref{fig:maxCC}   we illustrate this uncertainty by showing the maximal CC under the assumption of the prefactor above to lie anywhere between $0.5 < c_0 < 2$. Using Eq.~\eqref{eq:O4decayrate}, the corresponding rate can therefore be approximated as
\begin{equation}
\label{eq:O4decayrateapprox}
\frac{\Gamma_4(T) }{V} \sim  \frac{3 \pi^2}{4096}  (4-\epsilon_2)(3-\epsilon_2)N^4 T_*^4\ e^{-3 N^2/16} \,.
\end{equation}
As with $\Gamma_0^{(3)}$, here $\Gamma_0^{(4)}$ parametrizes our ignorance and following~\cite{Linde:1981zj} taken to be of order ${\cal O}(10^{-3}-10^3)$ and ${\cal O}(10^{-4}-10^4)$ respectively (see App.~\ref{app:uncertainties}).

\subsection{Dynamics of the  Phase Transition}
\label{sec:dynamicsPT}

\begin{figure}
\begin{center}
\includegraphics[width=14.5cm]{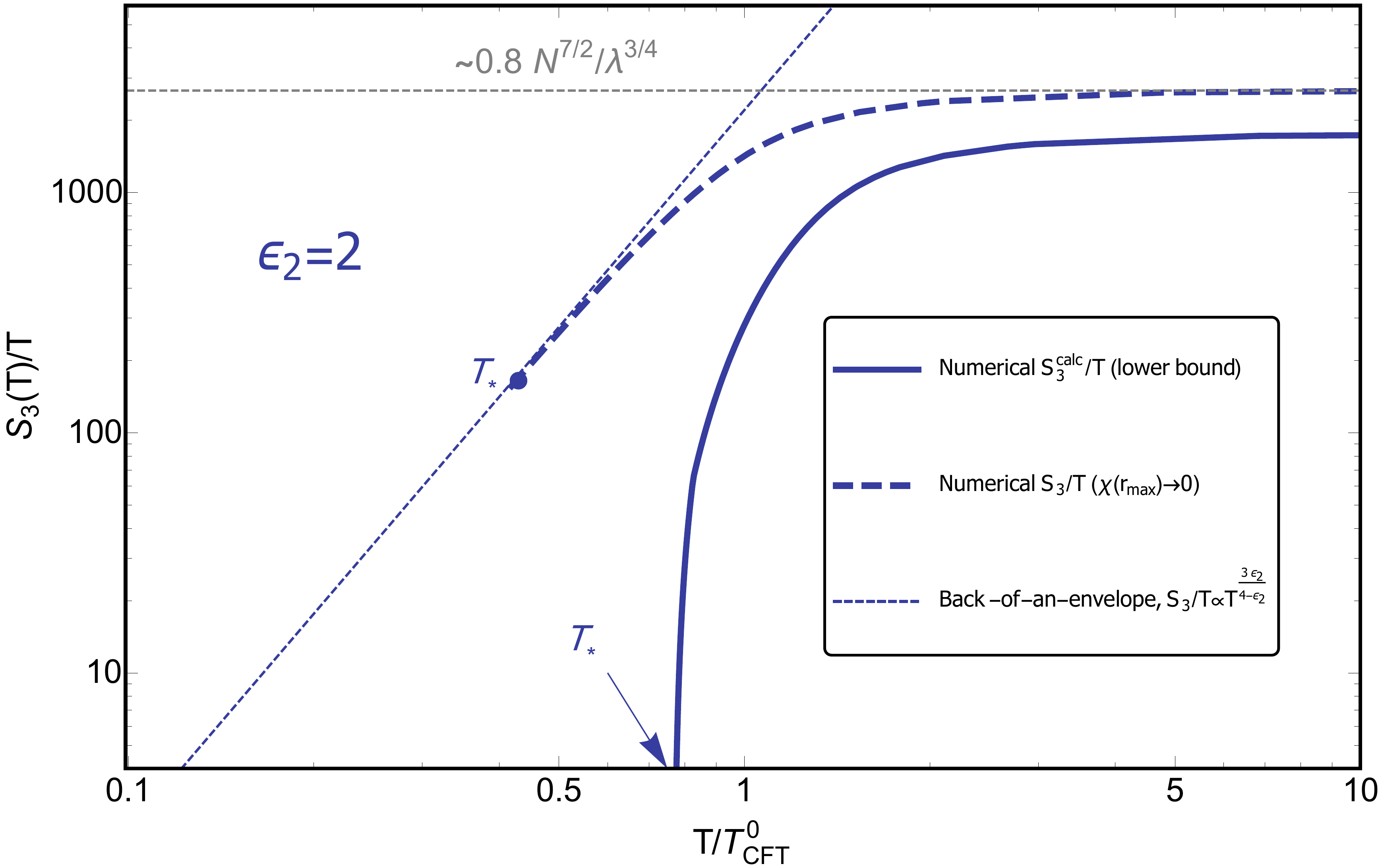}
\caption{Estimates of the thermal tunneling bounce action, $S_3(T)/T$, as a function of the temperature $T$, in units of the CFT temperature today, $T_{\rm CFT}^0$.  We compare three different estimates discussed in Sec.~\ref{sec:estimates} and App.~\ref{app:fullcalc}.  The {\bf solid line} shows our numerical evaluation of the calculable part of the $S_3(T)/T$ bounce action.   Since we discard the non-calculable contribution, this should be understood as a lower bound on the total action.  The corresponding temperature $T_*$, indicated by the arrow, should be viewed as a rough upper bound.  As discussed in the text, this value of $T_*$ agrees with our back-of-the-envelope estimate, Eq.~\eqref{eq:Tstar}.
The {\bf dashed line} is another numerical calculation, obtained by integrating the bounce action all the way to $\chi=0$, using the potential of Eq.~\eqref{eq:DilatonPotential}.   The corresponding $T_*$ is shown next to that line.   We expect the physical $T_*$ to lie around the above two values.  
The {\bf thin dashed line} shows the behavior of the rates using our back-of-the-envelope analytical understanding (See Eq.~\eqref{eq:S3estimate}).   These results show that at around $T_*$ the bounce action rapidly falls (and correspondingly the decay rate skyrockets) as expected.  
For this plot we fixed $\epsilon_2=2$, $\lambda = 1/30$ and $N=5$.   Furthermore, we choose $\chi_*$ to be such that $\left.S_3(T)/T\right|_{T=T^0_{\rm CFT}}=280$ for the numerical calculation depicted by the solid line.
\label{fig:action_1}} 
\end{center}
\end{figure}

We are now ready to understand the dynamics of the phase transition.  The rate   benefits from both the $O(3)$- and $O(4)$-symmetric contributions, each dominating at different temperatures with the transition taking place at  $T =1/R_{\rm bubble}$.   Using Eqs.~\eqref{eq:Rbubble} and~\eqref{eq:Tstar}, one finds $R_{\rm bubble} T_* \sim {\cal O}(1)$ and therefore   at around $T=T_*$ the $S_4$ action comes to dominate.   We arrive at the conclusion that the phase transition is guaranteed to complete near $T_*$  via the $O(4)$ bounce, assuming a sufficiently small expansion rate.   This condition, together with our estimates in Eqs.~\eqref{eq:S4envelope} and~\eqref{eq:O4decayrate}, are used in the next section to derive the maximal CC value for which the crunching mechanism works. 

At temperatures above $T_*$  the $O(3)$ bounce dominates, and the bounce action is calculable and large. The corresponding rate is falling as we dial up the temperature until it becomes comparable to $H_0$ in our universe. We choose the parameters in our model, and specifically $\chi_*$, so that this temperature corresponds to the temperature of the CFT in our universe today, $T_{\rm CFT}^0$ [for the approximate relation see Eqs.~\eqref{eq:decayrate},~\eqref{eq:Tstar} and~\eqref{eq:S3estimate}]. This ensures that our universe has not decayed yet.

The above behavior of the action is shown in Fig.~\ref{fig:action_1} for the case of $\epsilon_2=2$. The thick solid line corresponds to the numerical calculation (outlined in App.~\ref{app:fullcalc}) of the calculable part of the action only, thereby serving as a lower bound for the full action. The $T_*$ calculated with this method agrees with our estimation in Eq.~\eqref{eq:Tstar}. This is 
because Eq.~\eqref{eq:Tstar} is derived by neglecting the friction term in the EOM while the numerical calculation, which ignores the non-calculable regime, is equivalent to taking that part of the bounce action to be frictionless.  At $T=T_*$, only  the non-calculable regime contributes and hence the two methods coincide.  
 The thick dashed line corresponds to a similar  numerical  calculation but carried instead all the way to $\chi=0$, and thus including a region where such a calculation is invalid. This is a relatively good approximation for sizable $\epsilon_2$ and when $T\gg T_*$.  Under those conditions, the $T_*$ calculated this way varies by ${\cal O}(1)$ from the previous numerical calculation.  Conversely for $\epsilon_2\ll1$ one expects this calculation to result with a lower bound for $T_*$.  To
see this, we note that for $\epsilon_2\rightarrow 0$ the friction term can be shown (by simply solving the EOM) to diverge, requiring $\chi_r\rightarrow \infty$ and consequently $T_*\rightarrow 0$.    
Finally, the thin dashed line corresponds to the back of the envelope estimate outlined above. We see that this fits well to the calculation up to $\chi=0$ (thick dashed), when appropriately scaled by an ${\cal O}(1)$ factor which cannot be derived from the dimensional analysis. We conclude that the behavior of all three estimations is similar (for $\epsilon_2\sim {\cal O}(1)$), showing that  at temperatures close to $T_*$ the $O(3)$-symmetric bounce action drops rapidly.  Below we use the two different estimates for $T_*$ to establish a rough uncertainty band for our predictions.

As a final comment, we would like to reiterate that when we dial down $\epsilon_2$, the dependence of the bounce action on $T$ above $T_*$, $S_3(T)/T$,  becomes weaker.   Consequently, for a fixed $T_*$ [or roughly equivalently, fixed $\chi_*$ - see Eq.~\eqref{eq:Tstar}], the decay rate reaches values consistent with the stability of our observable universe only at significantly higher temperatures.   In other words, $T_*/T^0_{\rm CFT}$ drops together with $\epsilon_2$ thereby postponing the crunching process to low temperatures and implying a lower decay rate [which is proportional to $T_*^4$ - see Eq.~\eqref{eq:O4decayrateapprox}].   
As we show next, the faster the decay rate, the larger the CC which can be crunched away.  The success of this model therefore relies on sizable $\epsilon_2$ and as discussed earlier, this can be obtained with the introduction of an additional 
(QCD-like) asymptotically free gauge theory which explicitly breaks the CFT by order one at $\chi\simeq \chi_*$.

\section{Maximal CC}
\label{sec:CCmax}

\begin{figure}
\begin{center}
\includegraphics[width=7.5cm]{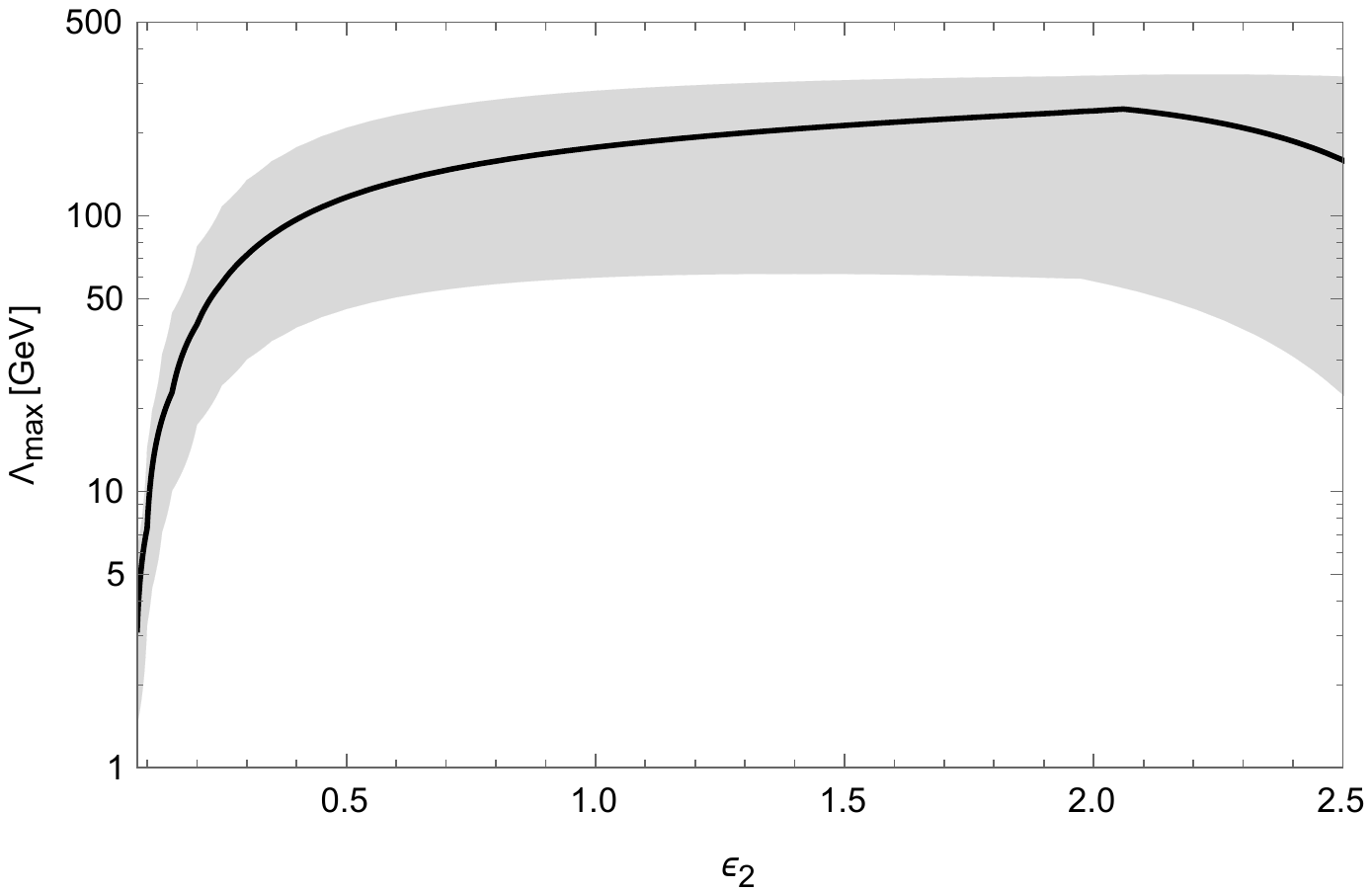}
\includegraphics[width=7.5cm]{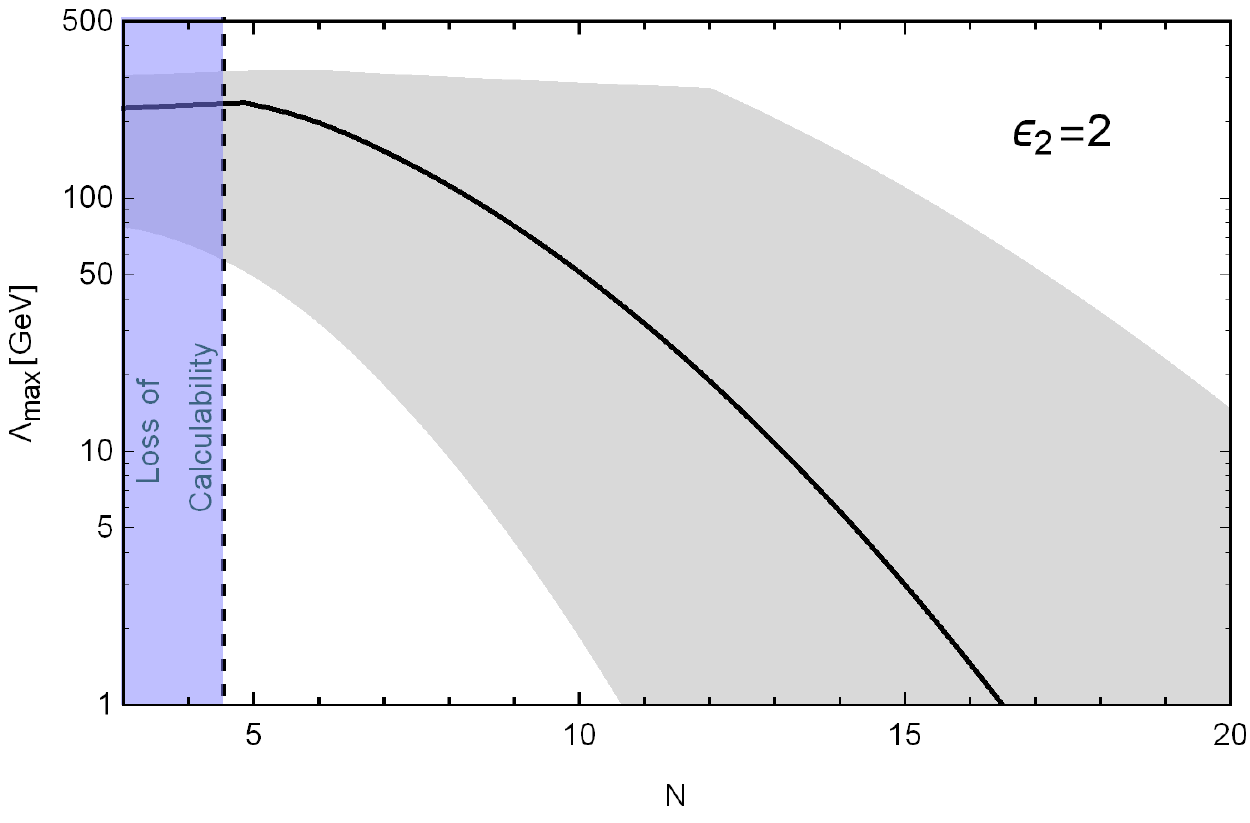}
\caption{
\label{fig:maxCC}
An estimate for the highest value of the CC, $\Lambda_{\max}$, below which the CFT sector dynamically triggers the crunching of the corresponding region of space, shown with a black solid line as a function of $\epsilon_2$ ({\bf left}) and the number of colors in the CFT, $N$ ({\bf right}).   $\Lambda_{\max}$ satisfies the constraints discussed in Sec.~\ref{sec:CCmax} and is calculated using our numerical methods laid out in App.~\ref{app:fullcalc}.
The gray region shows an uncertainty band which aims to take into account our ignorance when estimating the decay rate.   In particular, we vary $T_*$ in accordance to our two numerical estimates discussed in Sec.~\ref{sec:dynamicsPT} and illustrated in Fig.~\ref{fig:action_1}, and take   $\Gamma_0^{(4)}$ to be in the range $10^{-4}-10^4$ while  $S_4=(0.5-2)\cdot 3N^2/16$.  See App.~\ref{app:uncertainties} for more details.
For the left  plot we fix $\lambda=1/100$ and take $N$ to be the minimal value allowed by perturbativity and by our requirement that    quantum gravity corrections do not start dominating before $T_*$ [in accordance with Eq.~\eqref{eq:Nmin}]. 
For the right plot we fix $\epsilon_2=2$, and $\lambda=\frac{1}{60}(N/4.47)^{14/3}$. 
In that plot, the blue vertical region lies beyond our calculable regime.}  
\end{center}
\end{figure}

Having estimated the bounce action, we now turn to find the maximal CC, $\Lambda_{\rm max}$, which can be crunched away using our mechanism. 
The main limitation arises from the condition of completing the phase transition.  If one finds the decay rate, which peaks at around $T_*$ and is dominated by the O(4)-invariant bubble, to be small with respect to a region's Hubble rate, that region of space will never complete the crunching transition and we will be left with an eternally inflating universe.

Recall that we assume that the CFT sector is reheated at the end of inflation, and its temperature is initially above the critical temperature $T_c$. It is therefore in the symmetric phase of the CFT after reheating and remains there until $T\sim T_*$ at which point the decay rate suddenly increases to its maximal value. 
To ensure stability of our own patch, we require that around us the temperature of the CFT is above $T_*$. Other patches, specifically those with larger CC values, re-enter inflation early on and their  temperature drops exponentially until reaching $T\sim T_*$. At this point, the CFT transitions into the broken phase, contributing a large negative CC to  the energy density and subsequently causing the patch to crunch. 

Three  constraints determine $\Lambda_{\rm max}$:
\begin{enumerate}
\item{\bf Our patch should survive until today.}  This requirement translates to
\begin{equation}
\label{eq:maxCCconstraintToday}
\left. \frac{\Gamma}{V}\right|_{T_{\rm CFT}\geq T_{\rm CFT}^0}<H_0^4\,,
\end{equation}
and ensures that in our observable Hubble patch the CFT is still in the hot unbroken phase.
Since the $O(3)$-invariant bubble nucleation  dominates, this constraint can be rewritten using~\eqref{eq:decayrate} (see also App.~\ref{app:bouncetoday}) as,
\begin{equation}
\label{eq:s3today}
\left. \frac{S_3(T)}{T}\right|_{T_{\rm CFT}=T_{\rm CFT}^0}\gsim 280\,.
\end{equation}
In particular, within our construction this means that the nucleation temperature, which is around $T_*$, must be below the current temperature of the CFT, $T_{\rm CFT}^0$.   By choosing $T_{\rm CFT}^0$ such that the bounce action saturates the lower limit, Eq.~\eqref{eq:s3today},  one obtains, for a given $\epsilon_2$ and $\lambda$, a precise relation between the CFT's temperature and $T_*$ [see Eq.~\eqref{eq:S3estimate}].

\item{\bf $\mathbf{N_{\rm eff}}$.}  Since the CFT in our Hubble patch is in the hot conformal phase, it contributes to the effective number of relativistic degrees of freedom.  As a consequence, it is constrained by measurements of $N_{\rm eff}$~\cite{Aghanim:2018eyx} which place an upper bound on its temperature.   As we show in App.~\ref{app:bouncetoday},  the  95\%~C.L.~bound on the relativistic degrees of freedom implies the upper bound on the CFT temperature 
\begin{equation}
\label{eq:TCFT}
T_{\rm CFT}^0 \leq 0.034~{\rm meV} \left(\frac{N}{4.5}\right)^{-1/2}  \,.
\end{equation}
A lower bound on $N$ from naive dimensional analysis and also requiring $T_*$ to be larger than the temperature at which quantum gravity becomes important, is $N\gtrsim 4.47$~\cite{Agashe:2007zd} (see also App.~\ref{app:numericbounce}). Saturating  the above limit implies that the SM temperature, $T_0$, must be larger than the CFT temperature today, $T_{\rm CFT}^0$.  Thus together with 
the first condition we have:  $T_* \lesssim T_{\rm CFT}^0 \lesssim T_0$.

\item{\bf No eternal inflation.}  One of the main virtues of our solution is that no eternal inflation is needed to populate the variety of regions with varying CC.    An important constraint is therefore to ensure that the secondary phase of inflation, which begins once the CC in a given patch starts to dominate, always ends via the coalescence of true-vacuum bubbles.    In de-Sitter space, the temperature never drops below the Hawking-Gibbons temperature of order $H_{\Lambda} \simeq \Lambda^2/\sqrt{3} \bar M_{\rm Pl}$ (where $\Lambda$ is the CC in a given patch), and thus this condition can be written as,
\begin{equation}
\label{eq:maxCCconstraintNoEternal}
\left.\frac{\Gamma}{V}\right|_{T_{\rm CFT} = H_{\Lambda}} > H_\Lambda^4\,.
\end{equation}
When $H_\Lambda \ll T_*$, the above decay rate is dominated by the $O(4)$-symmetric bounce action and using Eqs.~\eqref{eq:S4envelope} and \eqref{eq:O4decayrateapprox}, we find,
\begin{equation}
\label{eq:LambdaMax}
\Lambda \lesssim \Lambda_{\rm max} \equiv \sqrt{N T_* \bar M_{\rm Pl}}\ e^{-3N^2 /128} \,.
\end{equation}
In the opposite limit, $H_\Lambda \gtrsim T_*$, the curvature of space, $H_\Lambda^{-1}$, is non-negligible, introducing  another dimensionful parameter to the theory and thereby precluding our ability to estimate $S_4$ on dimensional grounds.   Consequently, $H_\Lambda \lesssim T_*$ is taken as an independent constraint in our theory and plays a role when $T_*$ is small which occurs for $\epsilon_2\ll 1$. We additionally check that the classical rolling is dominant over quantum fluctuations at the release point, which is satisfied for $H_\Lambda \gtrsim \chi_*$.

\end{enumerate}

Putting all the constraints together, one arrives at the maximal CC that can be canceled,
\begin{equation}
\label{eq:LambdaMaxApprox}
\Lambda_{\rm max} \lesssim \sqrt{T_0 \bar M_{\rm Pl}}\ \sqrt{N}e^{-3N^2 /128} \lesssim  1.2 \units{TeV} \,,
\end{equation}
where in the last inequality we used $T_0 \simeq 2.73\units{kelvin}$ and the lower bound on $N$, $N\gtrsim 4.47$.
A more accurate maximal value of the CC, using the correct numerical factors and imposing the above constraints is shown on Fig.~\ref{fig:maxCC} as a function of $\epsilon_2$ (left) and $N$ (right).   In those plots the numerical bounce action is used and systematic  uncertainties discussed in App.~\ref{app:uncertainties} are shown.    We can see that for reasonable choices of the parameters, CCs slightly below one TeV may be crunched away.
 This scale $\Lambda_{\rm max}$, displayed in Fig.~\ref{fig:maxCC}, is where the appearance of new physics would be expected.   Remarkably, this scale is nothing more than the geometrical mean of $T_0$ and $\bar M_{\rm Pl}$, giving rise to the weak scale.  This ``CC miracle" is reminiscent of the well-known WIMP miracle,   providing  a whole new argument for new physics at the weak scale, unrelated to the Higgs hierarchy problem.

\section{Phenomenological Implications}
\label{sec:pheno}

\begin{figure}
\begin{center}
\includegraphics[width=10cm]{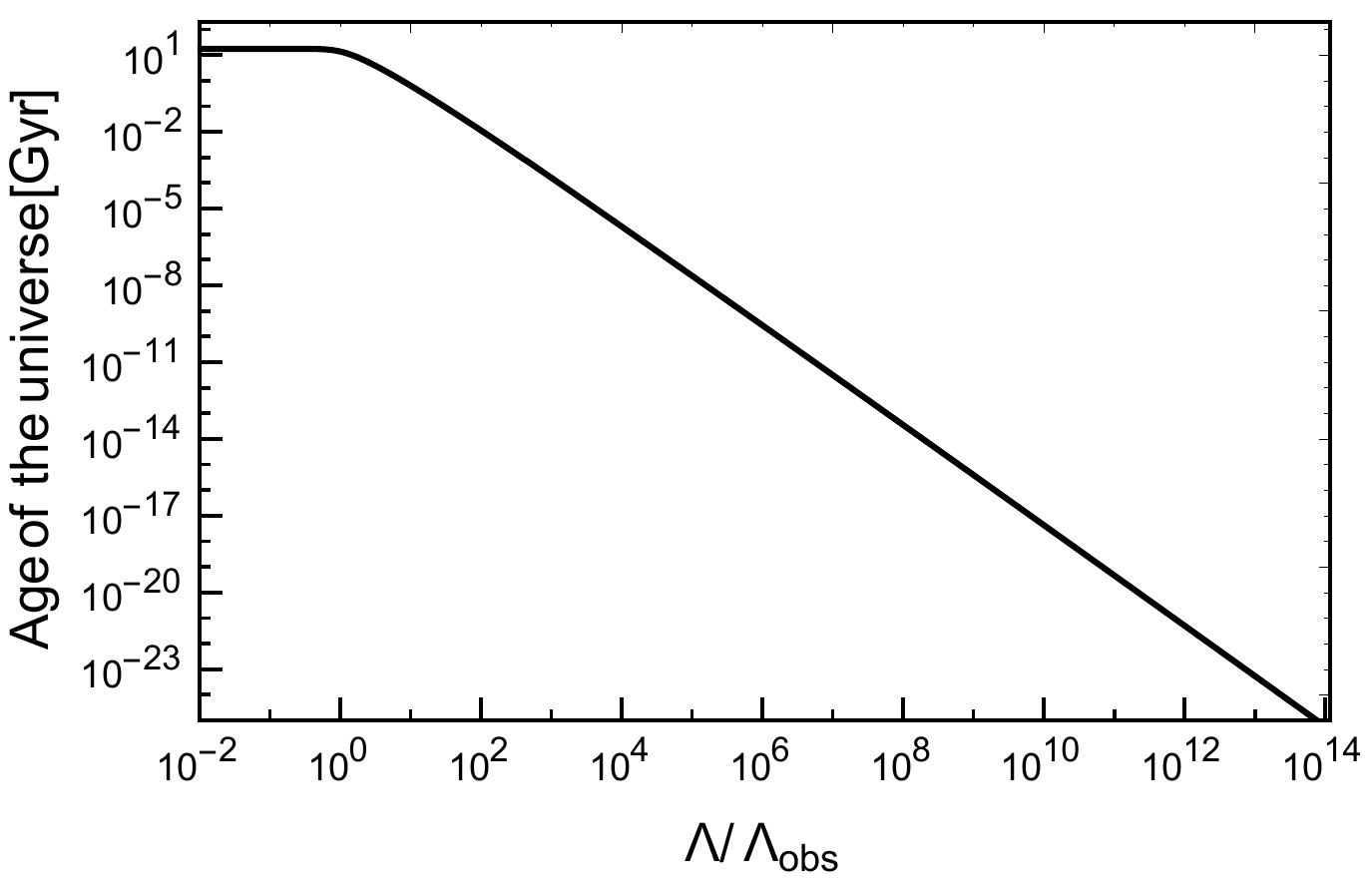}
\caption{The age of a given patch as a function of the CC value in it (in units of the observed CC).     For this plot we take $N=5$, $\epsilon_2=2$, and $\lambda=1/30$, which leads to $T_*\simeq 0.8 T_{\rm CFT}^{(0)}$, with other parameters at their nominal values (see App.~\ref{app:uncertainties}). 
The plot shows that as time passes, only patches with increasingly smaller values of the CC  survive.  For the parameters taken here, by $t\simeq 17\units{Gyr}$ all patches are predicted to crunch regardless of their CC value. \label{fig:ageCC}}
\end{center}
\end{figure}

The essence of our proposal is a mechanism that by today destroys every patch of the universe with a $\Lambda$ higher than the measured value in our observable domain. The timescale for this crunch is ${\cal O}\left(\Lambda^2/\bar M_{\rm Pl}\right)^{-1}$, i.e.~shorter than the age of the universe for such $\Lambda$. Therefore, the mechanism implies a relation, shown in Fig.~\ref{fig:ageCC},  between the maximal CC one might measure, and the time elapsed since the Big Bang. Therefore, since our universe is  old,  we should  expect to measure a small value of the CC, as all patches with values larger than the observable one have decayed already.

While our idea shares many similarities with the anthropic solution, there are significant conceptual differences. First, the mechanism avoids eternal inflation and its measure problem (assuming that it can be completed into a theory of non-eternal inflation, which seems entirely plausible; see e.g.~\cite{Mukhanov:2014uwa}).  Second, it has rather sharp experimental implications, contrary to the original landscape solution.  Here we shortly mention some of those, postponing a more detailed exploration to future work:
 \begin{enumerate}
 \item {\bf A measurable $N_{\rm eff}$.}  The hot  CFT  contributes to $N_{\rm eff}$, resulting in the bound on the temperature in that sector, Eq.~\eqref{eq:TCFT}.  Meanwhile, $T_{\rm CFT}^0$ is related to $T_*$ through the requirement that our Hubble patch survived till today, and therefore to $\Lambda_{\rm max}$ due to Eq.~\eqref{eq:LambdaMaxApprox} and the relation $T_{\rm CFT}^0 < T_0$.  As a consequence, one is led to a prediction of the form,
\begin{equation}
\Delta N_{\rm eff} \simeq 0.23 \left(\frac{\Lambda_{\rm max}}{260 \units{GeV}}\right)^8\,,
\end{equation}
where $\Delta N_{\rm eff} = N_{\rm eff} - N_{\rm eff}^{\rm SM}$, with $N_{\rm eff}^{\rm SM} = 3.046$. Given the current 95\% C.L.~limits, $\Delta N_{\rm eff} < 0.23$~\cite{Aghanim:2018eyx},  the value for $\Lambda_{\rm max}$ is in the 100-300 GeV range.

\item {\bf New physics at the weak scale.}  Our mechanism predicts the scale where new physics has to appear in order to cancel the UV contributions to the CC above $\Lambda_{\rm max}$. Remarkably  this scale turns out to be around the EW scale (since the highest value for $\Lambda_{\rm max}$ is around  $\sim300\units{GeV}$), independently of any considerations related to electroweak symmetry breaking, the Higgs hierarchy or the WIMP miracle.    An example of the new physics that could cancel the UV contributions to $\Lambda$ is  SUSY with a low breaking scale.

\item {\bf Our patch is about to crunch.}   We hope this prediction will not be verified anytime soon. 
 
\end{enumerate}

Barring additional modifications to the theory, the $N_{\rm eff}$ constraint together with the  absence of new physics at the LHC, already implies some mild tension for our mechanism.  Upcoming measurements for $N_{\rm eff}$ in CMB-S4 are therefore expected to find deviations from the SM prediction.  Conversely, if this constraint  becomes more severe with future measurements, the tension will grow.  
For instance, assuming that no new physics has been found  up to $\sim 2$ TeV, we can calculate the amount of tuning necessary, 
\begin{equation}
\frac{1}{\Delta} \sim \left( \frac{\Lambda_{\rm max}}{M_{NP}}\right)^4 \sim 0.03 \% ~\sqrt{\frac{0.23}{\Delta N_{\rm eff} }} \,.
\end{equation}
This is reminiscent of the Little Hierarchy   problem for the Higgs, and sub-percent fine tuning seems necessary for this realization of our paradigm.

\section{The Scanning and Inflationary Sectors}
\label{sec:scanning}
Before concluding, let us shortly discuss a requirement on the number of e-folds during the primary inflationary epoch. 

The existence of a landscape of vacua, i.e.~different values for the CC in different Hubble patches of the universe, plays a crucial role in our mechanism.  In this section  we wish to  show that the landscape can be populated without eternal inflation. 
Various mechanisms for  the scanning of the CC have been identified and studied extensively in the context of the anthropic approach (see e.g.~\cite{Brown:1987dd, Brown:1988kg, Bousso:2000xa,Martin:2013tda,Arvanitaki:2009fg, Bachlechner:2017zpb}). Here we will not revisit  these mechanisms, nor will we discuss the precise model of (non-eternal) inflation, which we assume to exist.  Instead, we limit our discussion to showing that with a finite and relatively small number of e-folds, a sufficiently dense population of the landscape may be produced.   

In our discussion we  think of the landscape as a large collection of vacua separated by barriers, and with a fixed value for the 
 tunneling rate $\Gamma_{\rm land}/V$. Other scenarios of rolling fields may be envisioned.   Moreover, one may  relax the constant decay rate assumption without changing the conclusions, as our results only depend logarithmically on the rate.
To ensure that the vacuum in our observable patch remains stable, the only  constraint on the rate reads
\begin{equation}
\frac{\Gamma_{\rm land}}{V} < H^4_0\,,
\label{eq:ourpatch}
\end{equation}
where $H_0$ is the Hubble constant today.  

There is a lower bound on the number of e-folds needed in order to populate the universe with a multitude of different values of the CC.
 To calculate it, we  separate inflation into two periods: the last $\sim$ 60 e-folds of inflation which correspond to the period where a single Hubble patch gets stretched to become our observable universe today,  and the first ${\cal N}$ that precede them. To calculate the number of vacua populated throughout the first ${\cal N}$ e-folds, we assume we start with a single Hubble patch of radius $1/H_{\rm inf}$ at the beginning of inflation. Inflation lasts ${\cal N}/H_{\rm inf}$, and during this time our initial volume grows by a factor $N_{\rm patch} = e^{3 {\cal N}}$. Hence the average number of decays, and consequently number of different CC values in our universe, is roughly given by
\begin{equation}
\langle N_{\rm dec} \rangle = N_{\rm patch} P_{\rm dec} \simeq e^{3 {\cal N}}\frac{\Gamma_{\rm land}}{V}\frac{{\cal N}}{H_{\rm inf}^4}   <  e^{3 {\cal N}}  {\cal N} \frac{H^4_0}{H^4_{\rm inf}}\,,
\label{eq:numberofpatches}
\end{equation}
where $P_{\rm dec} = [1- \exp(-\Gamma_{\rm land} t / V H_{\rm inf}^3)]\simeq \Gamma_{\rm land} {\cal N} / V H_{\rm inf}^4$ is the probability of a single patch to decay in ${\cal N}$ Hubble times, and we used (\ref{eq:ourpatch}) for the last inequality. 

A viable model of the landscape offers a sufficiently dense set of minima.   In order to find in such a model some patches with a CC as low as ours, $\Lambda_{\rm obs} \simeq 2\times 10^{-3}\units{eV}$, we need to make sure that the number of populated patches cover everything between $\Lambda_{\rm obs}$ and $\Lambda_{\rm max}$,
\begin{equation}
N_{\rm dec} > \frac{\Lambda^4_{\rm max}}{\Lambda^4_{\rm obs}} \simeq 4\times 10^{58}\left(\frac{\Lambda_{\rm max}}{\mathrm{TeV}}\right)^4\,.
\end{equation}
Using (\ref{eq:numberofpatches}) we find the minimal number of e-folds to be
\begin{equation}
 {\cal N} \gtrsim 133 + 1.3 \log\left(\frac{\Lambda_{\rm max} \Lambda_{\rm inf}^2}{\mathrm{TeV}^3}\right)\,,
\end{equation}
where $\Lambda_{\rm inf} \simeq \sqrt{H_{\rm inf}\bar M_{\rm Pl}}$ is the scale during inflation.
 As mentioned above, through Eq.~\eqref{eq:ourpatch}, this lower bound depends only logarithmically on the tunneling rate $\Gamma_{\rm land}/V$. Thus, we conclude  that a relatively short period of inflation can be sufficient to populate the landscape and scan over the CC at the level needed for our model.

\section{Discussion and Outlook} 
\label{sec:outlook}

In this paper we have proposed a new approach to addressing the CC problem. Much like the anthropic solution, a landscape of different vacua is assumed and the measured value of the CC around us has no special properties within it. However the paradigm discussed in this paper requires no eternal inflation and the mechanism for the selection of the CC is novel and introduces    a new form of cosmological dynamics. 

Our scenario  assumes 
that at zero temperature the landscape covers a range of strictly negative (and large) vacuum energies.   During the thermal history of the universe and after reheating, a hidden sector is driven to  a metastable vacuum, which raises the vacuum energies of the landscape to stretch from a large negative to a large positive CC. The average time spent in  the metastable vacuum within a given patch  depends on the local thermal history, which in turn depends on the value of the vacuum energy in that given patch. In particular, this time is longest when the CC is extremely small, so that the only patches that survive in this phase all the way  to the current age of the universe, have CC values smaller or comparable to the one measured in our patch. Once the metastable state decays in a given patch, the negative dark energy  dominates, causing the patch to crunch.

This metastable state was realized in this paper as the hot conformally symmetric phase of a spontaneously broken CFT.  This phase is known to be super-cooled, which postpones the phase transition to a very low temperature. The associated lifetime of this phase therefore depends on how long it takes the temperature to fall to the nucleation temperature associated with the phase transition. Any patch  with a large CC re-enters dark energy domination early, leading to an exponential drop of the temperature. The lifetime of the patch is therefore determined by the onset of dark energy domination, shortly after which the patch  crunches. Since a very small value of the CC is needed to delay dark energy domination to times comparable to the age of our universe, the measured value of the CC is not surprising, nor is it surprising that our vacuum energy only recently came to dominate;  had it done so much earlier, our observable universe would not have survived that long.  It is important to note that the bleak fate of vacuum decay is also predicted for our future.

The main advantage of this approach is twofold.  First, the finite lifetime of  any positive vacuum energy allows us to avoid eternal inflation. This means that the validity of our approach is not reliant on unambiguously defining a probability measure in eternal inflation~\cite{Linde:2010xz}, which is yet to be achieved. The requirement that the phase transition completes (thereby evading eternal inflation),
turns out to be very restrictive for our model  leading us to the second major advantage - predictivity. This condition together with  $N_{\rm eff}$ constraints on the radiation energy density  in the CFT sector, places an upper bound on the maximal value of the CC in the landscape that may be addressed using our mechanism.  For theories with a higher cutoff, distinct new physics (such as supersymmetry) must appear  to remove UV contributions from the CC. Miraculously this scale turns out to be around the weak scale, which arises as the geometric mean of $T_0$ and $M_{\rm Pl}$ and without any connection to the Higgs hierarchy problem.  
 Moreover, the lack of  new physics at the LHC already requires a little hierarchy between the weak scale and the landscape scale.   
 Finally, a second related prediction is an excess   in $N_{\rm eff}$ in future CMB experiments.

Future prospects for this idea may focus on a construction of more realizations which can stand in for the broken CFT in this paper. Our own realization did not allow for a full calculation of the phase transition rates, which we circumvented by  introducing fudge factors with associated uncertainties and using dimensional analysis estimates where all else failed. It is possible that different  models will allow for a straightforward calculation of the rates and may give rise to new experimental predictions in colliders or in the sky. Additionally, it is interesting to consider concrete UV-complete models that would allow for the cutoff to be taken parametrically larger than the weak scale, and potentially explain the little hierarchy required in our realization. 

Before closing, we would like to comment on the anthropic aspect of this idea.   Even though our solution predicts our universe to be eventually dominated by regions with small CC, the presence of observers is never mentioned.  It is therefore tempting to think that this solution is not anthropic.  However, since we never explain why our universe is so old, some (albeit weak) anthropic reasoning is needed.  Curiously, in the spirit of the Doomsday argument, we 
expect our universe to soon be destroyed and indeed our model predicts our demolition to be right around the corner.

 \section*{Acknowledgments}
We thank  Prateek Agrawal, Nima Arkani-Hamed, Brando Bellazzini, Tim Cohen, Matthew McCullough,  Raffaele D'Agnolo, Sergey Dubovsky, Gregory Gabadadze,  Nissan Itzhaki, Ami Katz, Matt Kleban, Simon Knapen, Patrick Meade, Alex Pomarol,  Diego Redigolo,  Josh Ruderman, David Shih, Raman Sundrum, Ken Van Tilburg, and Jesse Thaler for useful discussions. 
IB is grateful for the support of the Alexander Zaks Scholarship, The Buchmann Scholarship and of the Azrieli Foundation.  IB and TV  thank the Erwin Schr\"odinger International Institute for hospitality while this work was in progress.
CC thanks the Aspen Center for Physics, the KITP at Santa Barbara as well as the Theoretical Physics Group at TU Munich  for their hospitality while this work was in progress. The research of CC was supported in part by the NSF grant PHY-1719877. CC and TV were supported in part by the BSF grant 2016153. 
MG thanks KITP at Santa Barbara and MIAPP for hospitality while this work was in progress. The research of MG was supported in part by the Israel Science Foundation (Grant No. 1302/19) and by the United States-Israel Binational Science Foundation (BSF) (Grant No. 2018236).
TV is further supported in part by 
the Israel Science Foundation-NSFC (grant No. 2522/17), by the European Research Council (ERC) under the EU Horizon 2020 Programme (ERC- CoG-2015 - Proposal n. 682676 LDMThExp),
and by a grant from the Ambrose Monell Foundation, given
by the Institute for Advanced Study.

\appendix

\section{Goldberger-Wise Stabilization of the RS Model\label{app:GW}}

In this Appendix we review the essential aspects of the GW stabilization mechanism for warped extra dimensions needed for the discussion of our model. As explained in Sec.~\ref{sec:crunching}, we start with the RS warped 5D theory~\cite{Randall:1999ee} with metric (\ref{eq:AdS}) and add a bulk scalar field $\phi$ to stabilize~\cite{Goldberger:1999uk} the dilaton, which in the 5D picture is identified with the radion field~\cite{Csaki:1999mp}, setting the size of the extra dimension. The full action of the RS-GW model is given by:
\begin{eqnarray}
S&=&\int d^4x dz  \sqrt{g} \left(g^{MN}\partial_M \phi \partial_N \phi + \Lambda_{\rm bulk}^5- m_{\rm bulk}^2\phi^2 \right) \nonumber \\
&& -\int d^4 x \sqrt{g_{\rm ind}(z_{\rm UV})} V_{\rm UV}(\phi(z_{\rm UV})) -\int d^4 x\sqrt{g_{\rm ind}(z_{\rm IR})} V_{\rm IR}(\phi(z_{\rm IR})))
\end{eqnarray}
where the localized potentials are 
\begin{eqnarray}
V_{\rm IR} &=& -\lambda_{\rm IR} \frac{v_{\rm IR}^2 }{z_{\rm IR}} \phi^2 +\lambda_{\rm IR} z_{\rm IR}^2 \phi^4 -\sqrt{\Lambda_{\rm bulk}^5 M^3_*}  + \delta_{\rm IR}  \nonumber \\
V_{\rm UV} &=& -\lambda_{\rm UV}  \frac{v_{\rm UV}^2}{z_{\rm UV}} \phi^2 +\lambda_{\rm UV} z_{\rm UV}^2 \phi^4 + \sqrt{\Lambda_{bulk}^5 M^3_*} + \delta_{\rm UV}  
\end{eqnarray}
The constant terms, i.e.~the brane tensions, are separated into the tuned piece $\pm \sqrt{\Lambda_{\rm bulk}^5    M^3_*}$ (the terms needed for a static solution without stabilization) and the detuning $\delta_{\rm UV, IR}$. The UV mistuning simply contributes to the 4D CC and does not play a role in the stabilization dynamics, up to a Hubble scale mass for the radion, which we neglect. 
The bulk solution to the EOMs, assuming a small bulk mass $m^2_{\rm bulk} = 4 \epsilon/z_{\rm UV}^2$, is given by  
\begin{equation}
\phi(z) = \frac{C_1}{z_{\rm UV}^{3/2}} z^{4+\epsilon} +\frac{C_2}{z_{\rm UV}^{3/2}} z^{-\epsilon}\,.
\end{equation}
Taking for convenience the limit of large $\lambda_{\rm IR},\lambda_{\rm UV}$, the boundary conditions are $\phi(z_{\rm UV})=\frac{v_{\rm UV}}{z_{\rm UV}^{3/2}}$ and $\phi(z_{\rm IR})=\frac{v_{\rm IR}}{z_{\rm IR}^{3/2}}$ and the full solution is:
\begin{equation}
C_1 \simeq v_{\rm IR}\left(\frac{z_{\rm UV}}{z_{\rm IR}} \right)^{4+\epsilon} + v_{\rm UV} \left(\frac{z_{\rm UV}}{z_{\rm IR}} \right)^{4+2\epsilon}~,~C_2 \simeq v_{\rm UV}\ .
\end{equation}
Integrating out $\phi$ by plugging this solution back into the action we find that the energy has the following dependence~\cite{Csaki:2000zn,Rattazzi:2000hs} on the inter-brane separation $\chi=1/z_{\rm IR}$ 
\begin{equation}
V(\chi) \simeq \left(\delta_{\rm IR}/k^4 +4v^2_{\rm IR}\right) \chi^4 - \frac{8}{k^{\epsilon}} v_{\rm IR}v_{\rm UV} \chi^{4+\epsilon}  + \frac{4 v^2_{\rm UV}}{k^{2\epsilon}} \chi^{4+2\epsilon} \equiv  
\lambda \chi ^4 - \lambda_{\epsilon}k^{-\epsilon} \chi^{4+\epsilon} + \lambda_{2\epsilon}k^{-2\epsilon}\chi^{4+2\epsilon}
\end{equation} 
where the $\lambda$'s are taken to be positive and $\epsilon$ negative\footnote{Note that once we introduce QCD, we will be interested with a positive $\epsilon$ instead.}. We will take $\lambda>\lambda_{\epsilon}>\lambda_{2\epsilon}$, while keeping all of them ${\cal O}(1)$. For values of $\chi$ satisfying 
\begin{equation}
\chi> \hat{\chi}= \left(\frac{\lambda_{2\epsilon}}{\lambda_{\epsilon}}\right)^{-\frac{1}{\epsilon}}k\,,
\end{equation}
the $\chi^{4+2\epsilon}$ term can be neglected. For small $\epsilon$, $\hat{\chi}$ can be arbitrarily small hence we can use the following expression for the dilaton potential of the RS-GW model:
\begin{equation}
V(\chi) \simeq \lambda \chi ^4 - \lambda_{\epsilon}k^{-\epsilon} \chi^{4+\epsilon}\label{eq:appendix_DilatonPotential}
\end{equation}
The $\chi$ scalar is identified via the AdS/CFT correspondence with the Goldstone boson of the spontaneous breaking of the conformal symmetry - the dilaton~\cite{Rattazzi:2000hs,ArkaniHamed:2000ds}.  On the CFT side RS-GW is a theory of a spontaneously broken conformal symmetry in the presence a near marginal deformation with dimension $\Delta=4+\epsilon$, explicitly breaking the conformal symmetry. The leading order dilaton potential given by Eq.~\eqref{eq:appendix_DilatonPotential}  accounts for the first two terms used in our model, Eq~\eqref{eq:DilatonPotential}.

\section{Effects of an Additional Bulk Gauge Group}
\label{app:QCD}

While the RS-GW model reviewed in App.~\ref{app:GW} produces the first two terms in (\ref{eq:DilatonPotential}), we have seen that an additional term is needed in order to enhance the bubble nucleation rate at low temperatures.  To achieve this we can use the recent proposal of von Harling and Servant~\cite{vonHarling:2017yew} who were investigating possible scenarios for facilitating the RS phase transition. They noticed that the additional scale dependence introduced into the RS model when the effects of QCD are taken into account, can significantly affect the nucleation probabilities of the broken phase at temperatures around the QCD scale (and well below the EWSB scale). Their assumption was that QCD lives in the bulk of the extra dimension (corresponding to a ``partially composite'' gluon field). Hence we will also assume the presence of an additional bulk $G=SU(N)$ gauge group (unrelated to QCD which is elementary in this picture and lives on the UV brane). 
The running of the coupling of $G$ will be given by~\cite{Csaki:2007ns} 
\begin{equation}
\frac{1}{g^2(Q, \chi)} = \frac{\log \frac{k}{\chi}}{k g_5^2} -\frac{b_{\rm UV}}{8\pi^2} \log \frac{k}{Q} -\frac{b_{\rm IR}}{8\pi^2} \log \frac{\chi}{Q} +\tau_{\rm UV}+\tau_{\rm IR} \,,\label{eq:runningcoupling}
\end{equation} 
where $Q$ is the running scale and the dependence on $\chi$ is introduced due to the finite size of the extra dimension. In this equation $k=1/z_{\rm UV}$ is the AdS curvature, $\tau_{\rm UV,IR}$ are the brane localized kinetic terms on the two branes and $b_{\rm UV,IR}$ are the 4D beta functions of the fields localized on the UV/IR branes. Note that (\ref{eq:runningcoupling}) is valid only for $Q \lesssim \chi$. Nominally for $Q>\chi$ the coupling will be independent of $\chi$. We will get back to this issue later. Using (\ref{eq:runningcoupling}) we can find the scale $\Lambda (\chi )$ where the coupling diverges~\cite{vonHarling:2017yew}:
\begin{equation}
\Lambda (\chi ) = \left( k^{b_{\rm UV}} \chi^{b_{\rm IR}} e^{-8\pi^2\tau} \left( \frac{\chi}{k}\right)^{-b_{\rm CFT}}\right)^\frac{1}{b_{\rm UV}+b_{\rm IR}} = \Lambda_0 \left( \frac{\chi}{\chi_{\min}} \right)^n \,,
\label{eq:Lambda}
\end{equation}
where
\begin{equation}
n = \frac{b_{\rm IR}-b_{\rm CFT}}{b_{\rm UV}+b_{\rm IR}}\,t .
\end{equation}
$\Lambda_0$ is the dynamical scale when the dilaton is at its minimum, $\chi_{\rm min}$, and $b_{\rm CFT}= -\frac{8\pi^2}{kg_5^2}$. The expression (\ref{eq:Lambda}) for $\Lambda (\chi )$ is valid only when $\chi \gsim \Lambda (\chi )$, corresponding to the fact that the scale of spontaneous  symmetry breaking happens before the condensation in this additional gauge group. The physics behind this requirement is to make sure that the explicit breaking introduced by the scale $\Lambda$ can be treated as a small perturbation on the approximately conformal theory. Otherwise the explicit breaking will be larger than the spontaneous breaking, and one should not be using the original dilaton $\chi$ as the low-energy degree of freedom. There may be another low-energy effective theory in terms of branes and another dilaton field, however its description strongly depends on the details of the dynamics. Hence the regime $\chi < \Lambda (\chi )$ is model dependent and not calculable within our framework. In our calculation for the tunneling probability we will have to clearly separate the contributions that can be reliably obtained from the calculable region, and provide estimates for the corrections from the non-calculable regime. 

The contribution of the extra dynamics to the dilaton potential is expected to be of the form 
\begin{equation} 
V_{G}=-\alpha \Lambda^4_{G} (\chi )\,,
\end{equation}
where $\alpha$ is an ${\cal O}(1)$ coefficient that depends on the exact matter content and dynamics of this new confining $SU(N)$ theory. For example \cite{vonHarling:2017yew} found that for the contribution of the gluon condensate in QCD, $\alpha = \frac{\beta_{\rm QCD}}{17}$.  Using (\ref{eq:Lambda}) we find that the form of the induced dilaton potential will be 
\begin{equation}
\label{eq:VQCD}
V_{G} = -\alpha \Lambda_0^4 \left( \frac{\chi}{\chi_{\min}}\right)^{4n} \,. 
 \end{equation} 
 From Eq.~\eqref{eq:VQCD}, we can see that $n$ is identified with  $(1-\epsilon_2/4)$ from Eq.~\eqref{eq:DilatonPotential}.

\section{Detailed Derivation of the Bubble Nucleation Rate}
\label{app:fullcalc}

In this appendix, we elaborate on the bubble nucleation-rate calculation presented in Sec.~\ref{sec:PT}. In App.~\ref{app:numericbounce} we present our methods to calculate   the $O(3)$-symmetric action, for a given set of  model's parameters. In App.~\ref{app:bouncetoday} we show how to choose the parameters of the theory using the constraints of Sec.~\ref{sec:CCmax}. We finish in App.~\ref{app:uncertainties}, where we describe in detail the various uncertainties in our calculations, and our estimates for them.

\subsection{Numerical Calculation of the Bounce Action\label{app:numericbounce}}

To find the O(3)-symmetric bounce action, $S_3$, we first need to find the O(3)-symmetric bounce solution. This bounce solution is described by the solution to the 1D EOM, Eq.~\eqref{eq:S3EOM}, with $\phi = \sqrt{3(N^2-1)/(2\pi^2)} \chi$ and $V(\chi)$ taken from Eq.~\eqref{eq:DilatonPotential}. However, the change in the field $\chi$ does not cover the entire route taken in field space during the tunneling (see Fig.~\ref{fig:Roads}). Therefore, our first step in this calculation is to understand exactly what is the region in which $\chi$ is the correct degree of freedom.

The second step is to find the dependence of the release point, $\chi_r\equiv \chi(r=0)$ on the temperature. The dependence of $\chi_r$ on the temperature is through the boundary conditions. To solve a second order differential equation, one needs two boundary conditions. The first boundary condition is $\chi'(0)=0$, while the second requires that the false vacuum is reached at $r\rightarrow \infty$. However, as $\chi$ is no longer a valid degree of freedom at $r\rightarrow \infty $, we need to find a method for estimating the effect of the non-calculable regime. Therefore this second boundary condition will be only approximately satisfied, sourcing an uncertainty on our calculation.
In practice, rather than imposing the second boundary condition, we solve the equation of motion for an arbitrary $\chi(0)=\chi_r$, and later use the second boundary condition to derive $\chi_r(T)$.

\subsubsection*{Finding the region of validity}
In Sec.~\ref{sec:PT}, we had two methods of calculating $S_3(T)/T$, one which only includes the calculable region and serves as a lower bound, and one that contains an estimate for the action in the non-calculable region as well. For the lower bound calculation we restrict our integration to the calculable region. The calculable region has to satisfy the following two conditions:
\begin{enumerate}
	\item $\chi\gsim T\, \max[(k /M_*),1]$. For smaller values of $\chi$,  additional KK modes are excited and/or the local Planck scale on the IR brane drops below the temperature, and quantum gravity corrections become important~\cite{Creminelli:2001th}.

	\item $\chi>\chi_*$. This condition ensures that the back-reaction due to the explicit breaking of the conformal symmetry is small~\cite{vonHarling:2017yew}.   Once violated, non-calculable corrections to the effective potential dominate and, equivalently, corrections to the AdS metric on the gravity side, which arise   from back-reaction,  are large.
\end{enumerate}

To derive the calculable contribution to the bounce action, one uses Eq.~\eqref{eq:S3} with the upper limit replaced by $r_{\rm max}$, which is the maximal $r$ for which $\chi$ reliably describes the theory. We stress that  this restriction of the integration region implies that our calculation should be viewed as a lower bound on the bounce action and therefore an upper bound on the  rate. This is conservative since $S_3(T)/T$ controls the decay  rate of our visible yet-to-decay unstable Hubble patch and since we require this rate to be sufficiently low to ensure our universe's survival till today.   We thus determine $r_{\rm max}$ as a function of $\chi_r$, $T$ and the parameters of the potential,  through the condition
\begin{equation}
\label{eq:Defrmax}
\bar\chi(r=r_{\rm max}; \chi_r)=\max\left[\chi_*,T,\left(\frac{16\pi^2}{N^2-1}\right)^{1/3} T\right] \, .
\end{equation}
We remind the reader that $\bar\chi(r; \chi_r)$ is the solution to the EOM which in turn depends on the boundary condition, $\chi_r$.  

At high temperatures, the first argument in the max of Eq.~\eqref{eq:Defrmax} is negligible to the other two. To simplify our discussion we will now impose a lower bound on $N$ that will ensure that as the temperature is lowered, the backreaction condition takes over at $T=T_*$ or above.

\subsubsection*{Comparing the Effects of Explicit Conformal Breaking  and High Temperature Effects}

When defining $T_*$ in Eq.~\eqref{eq:TstarDef} we were careful to evaluate the bounce solution over the calculable region where the backreaction of the metric is negligible by making sure $\chi > \chi_*$. However, as we discussed in this appendix, a theory of quantum gravity might also be needed in order to describe the region  $\chi<T\,k/M_*$ and additional KK states must be introduced for $\chi<T$. We thus define $\chi_{\rm QG}(T)\equiv  T\,\max(1,k/M_*)$ and act to ensure that $\chi_{\rm QG}(T_*)\leq \chi_*$, or else the theory at $T_*$ is already out of control.
To ensure this, we define the temperature $T_{\rm QG=BR}$ for which the $\chi_*$ yielding the regime of validity due to backreaction is the same as the $\chi_{\rm QG}$. For now let us assume that $k/M_*>1$, and $\chi_{\rm QG}(T)=Tk/M_*$, and we return to the case for $\chi_{\rm QG}=T$ below. We therefore define $T_{\rm QG=BR}$ by,
\begin{equation}
\frac{k}{M_*}T_{\rm QG=BR}=\chi_*~.
\end{equation}
In particular we should check that this temperature is higher than $T_*$, defined in Eq.~\eqref{eq:Tstar}. If that is indeed true, then $\chi_{\rm QG}(T_*)<\chi_{\rm QG}(T_{\rm QG=BR})=\chi_*$. Therefore, the condition we examine is,
\begin{equation}
\label{eq:NminCond}
\frac{T_{\rm QG=BR}}{T_*}=\frac{\left(\frac{N^2-1}{16\pi^2}\right)^{1/3}}{\left(\frac{12}{(4-\epsilon_2)(3-\epsilon_2)\pi^4}\right)^{1/4}}>1\,.
\end{equation}
We therefore restrict the region of $N$ (for a given $\epsilon_2$) to be
\begin{equation}
\label{eq:Nmin}
N\gsim2.26\left(\frac{12}{(4-\epsilon_2)(3-\epsilon_2)}\right)^{3/8}~.
\end{equation}
While the expression used for $T_*$ here was found using simple back of an envelope calculation, this minimal value of $N$ is only approximate, but the main point is that by restricting $N$ we can always ensure that the region of calculability is also within the regime of validity for the effective theory.

For $1>k/M_*$, increasing $N$ no longer reduces $\chi_{\rm QG}(T)$ for a fixed temperature. We thus cannot find  $N$ for which our theory is valid at $T=T_*$. By examining Eq.~\eqref{eq:NminCond}, we can see that when $\epsilon_2\gsim 2.9$, indeed this problem arises. We therefore restrict ourselves only to $\epsilon_2\leq 2.9$.

\subsubsection*{The Euclidean Energy}
$\chi_r$ and the temperature, $T$, are related through the boundary conditions of the EOM.  To see this, consider the Euclidean energy, the analog of energy for a particle in an inverted potential,
\begin{equation}
\label{eq:euce}
E_{\rm E}(r)=\frac{3(N^2-1)}{4\pi^2}\chi'(r)^2-V[\chi(r)]\,.
\end{equation}
One has  $E_{\rm E}(r\rightarrow 0)=-V_{\rm eff}(\chi_r)$ and   $E_{\rm E}(r\rightarrow \infty)=-V_{\rm CFT}\simeq -(\pi^2/8)N^2T^4$.   In the absence of friction, (the second term of Eq.~\eqref{eq:S3EOM}), the Euclidean energy is conserved and the relation between $\chi_r$ and $T$ follows trivially.   
With friction included, this energy is monotonically decreasing and using Eq.~\eqref{eq:S3EOM}, can be shown to obey,
\begin{equation}
\label{eq:eucer}
\frac{dE_{\rm E}}{dr}=-\frac{3(N^2-1)}{\pi^2}\frac{\chi'^2}{r}\,.
\end{equation}
In the BB region, the only scale is the temperature. Since the energy lost per unit distance is $\propto 1/r$,  Eq.~\eqref{eq:eucer} implies that the energy lost in the BB region would be $\propto T^3/(R_{\rm bubble})$. For $T> T_*$, we expect that region to be $\mathcal{O}(1)$ of the route taken in field space for $r>r_{\rm max}$. We therefore estimate
\begin{equation}
\label{eq:frictionestimate}
E_{\rm E}(r\rightarrow\infty)-E_{\rm E}(r_{\rm max}))\propto -\frac{1}{R_{\rm bubble}T}\cdot T^4\,.
\end{equation}
At $T\gg T_*$, using Eq.~\eqref{eq:Rbubble} and Eq.~\eqref{eq:chirT}, we see that $R_{\rm bubble}T\gg 1$, so we expect that $|E_{\rm E}(r\rightarrow\infty)-E_{\rm E}(r_{\rm max})|\ll T^4$. Since $E_{\rm E}(r\rightarrow\infty)=-V_{\rm CFT}(T)\propto T^4$, we can neglect this friction compared with the remaining energy. As T approaches $T_*$, we have $T\sim T_*\sim \chi_*\sim \chi_r$ and so all dimensionfull scales of the theory are of the same order. Therefore, we expect at that point that the lost energy would be $\sim T^4$ from dimensional considerations. Consequently we parametrize the energy lost to friction at any temperature $T>T_*$ after exiting the region of validity as
\begin{equation}
\label{eq:friction}
E_{\rm E}(r_{\max})+V_{\rm CFT}(T)=\alpha(T)T^4\,.
\end{equation}
The above equation relates the CFT potential which depends on $T$, to  $E_{\rm E}(r_{\rm max})$ which depends on $\chi_r$. As we have outlined above, we expect $\alpha(T)\ll 1$ at high temperatures, and $\alpha(T) \sim \mathcal{O}(1)$ at low temperatures.

In this discussion, we did not find the $N$ dependence of $\alpha$. Since we work in the large $N$ limit, this could significantly affect our estimate for $\alpha$ -- so the claims made in the  previous paragraph should be taken with a grain of salt.   For the (rather conservative) lower bound we take $\alpha(T)=0$.

\subsubsection*{Virialization of the Kinetic and Potential Terms}

The final additional ingredient is more of a technicality, and is only needed to increase the numerical accuracy of the integration over the Euclidean Lagrangian. Generally, derivatives of numerical functions are often calculated with a much lower precision than the original functions themselves, so that numerically integrating the kinetic term can introduce sizable numerical errors.   A mathematical trick which is often employed, is to integrate Eq.~\eqref{eq:S3} by parts, and use Eq.~\eqref{eq:S3EOM}, to relate the integral over the potential and the integral over the kinetic term. This  implies an effective ``virialization" of the kinetic and potential terms, for the entire integration region. The common use for this is done for the entire $[0,\infty)$ region of $r$, so that the boundary term at $r\rightarrow \infty$ is 0. However, for our integration region, $0$ to $r_{\max}$, the boundary term at $r_{\rm max}$ cannot be ignored, and one finds
\begin{equation}
\label{eq:scalc}
S_{3}^{\rm calc}(T)=4\pi \left[r_{\max}^3\left(\frac{2V_{\rm CFT}(T)}{3}-E_{\rm E}(r_{\max})\right)-2\int^{r_{\max}}_0 r^2 V_{\rm eff}[\chi(r)]dr\right]~.
\end{equation}
With this we are finally able to find the action $S_3(T)/T$ for any temperature which has a non-zero valid region. The final step in finding $S_3(T)/T$ is to add the non-calculable action to the calculable one, as discussed before.

\subsection{Evaluation of the Cosmological Constraints\label{app:bouncetoday}}

Having discussed the procedure for obtaining $S_3(T)/T$, we are now ready to present the details of the three cosmological constraints presented in Sec.~\ref{sec:CCmax}.  The first two conditions are required in order for our theory to be experimentally viable, while the third constraint ensures the phase transition completes in all patches of the universe, thereby evading eternal inflation.  Since we have already thoroughly discussed the third constraint in Sec.~\ref{sec:CCmax}, we do not discuss it here again. Instead, in this appendix we carefully examine the implications of the first two constraints, deriving Eqs.~\eqref{eq:s3today} and \eqref{eq:TCFT}.

\subsubsection*{Our Universe has Not Crunched Yet}

The first constraint in Sec.~\ref{sec:CCmax}, Eq.~\eqref{eq:maxCCconstraintToday}, is equivalent to a lower bound on the  bounce action at the current temperature of the CFT, $S_3(T_{\rm CFT}^0)/T_{\rm CFT}^0$. For lower values of the bounce action  our observable universe would not have survived till today.

Using  Eq.~\eqref{eq:decayrate}, the measured value of $H_0$~\cite{Aghanim:2018eyx}, and our upper limit for $T_{\rm CFT}^0$ in Eq.~\eqref{eq:TCFT}, we can derive the lower bound on $S_3(T)/T$. The explicit form of the bound involves the Lambert W function, $W_{-1}$,  which is the first negative solution for the equation, $z=W_{-1}(z)e^{W_{-1}(z)}$,

\begin{equation}
\frac{S_3(T_{\rm CFT}^{0})}{T_{\rm CFT}^{0}}=-3 W_{-1}\left(-4\pi H_0^{8/3}(\Gamma_0^{(3)})^{-2/3}(T_{\rm CFT}^0)^{-8/3}/3\right)/2\ .
\end{equation}
In the relevant regime, $W_{-1}$ is only logarithmically dependent upon its argument, so that the dependence on the exact values of $\Gamma_0^{(3)}$ and $T_{\rm CFT}^0$ is minimal. We thus find that for reasonable values of the parameters
\begin{equation}
\label{eq:stoday}
\frac{S_3(T_{\rm CFT}^0)}{T_{\rm CFT}^0}\gsim 267\pm 7.
\end{equation}
Since we neglect the logarithmic correction, we have chosen the more conservative  $S_3(T_{\rm CFT}^0)/T_{\rm CFT}^0=280$ in Eq.~\eqref{eq:s3today}.

\subsubsection*{The $N_{\rm eff}$ Constraint}

Next we focus on the second constraint Eq.~\eqref{eq:TCFT} discussed in Sec.~\ref{sec:CCmax}. 
During the times of BBN and CMB, the CFT  acts as a new contribution to the radiation energy density. Since the experimental observations  of the energy density from relativistic degrees of freedom during CMB and BBN~\cite{Aghanim:2018eyx} agree with the SM predictions, we will obtain a constraint  on the energy density of the CFT at those times. The experimental bounds are usually expressed in terms of a bound on the effective number of neutrinos ~\cite{Aghanim:2018eyx},
\begin{equation}
\label{eq:neff}
\rho_{\mathrm{rad}}=N_{\mathrm{eff}}\cdot \frac{7}{8}\left(\frac{4}{11}\right)^{4 / 3} \rho_{\gamma}\ ,
\end{equation}
where $\rho_\gamma=\pi^2T_\gamma^4/15$ is the energy density in ordinary photons, with $T_\gamma$ the  photon  temperature. The energy density of the CFT~\cite{Creminelli:2001th} is
\begin{equation}
\label{eq:eCFT}
\rho_{\rm CFT}=-3V_{\rm CFT}=\frac{3\pi^2}{8}N^2T_{\rm CFT}^4\ .
\end{equation}
The temperature of the CFT sector at any time $t$ (after BBN) obeys,
\begin{equation}
\label{eq:TCFTtoTSM}
T_{\rm CFT}(t)=\frac{T_{\gamma}(t)}{T_{\gamma}^0}\cdot T_{\rm CFT}^0 
\end{equation}
where the superscript 0 refers to the temperature measured today, and $T_{\gamma}$, $T_{\rm CFT}$ are the temperatures of ordinary photons and the hidden CFT respectively. By plugging Eq.~\eqref{eq:TCFTtoTSM} and Eq.~\eqref{eq:eCFT} in Eq.~\eqref{eq:neff}, we derive the 95\%~C.L. bound on the temperature of the CFT today,
\begin{equation}
T_{\rm CFT}^0\leq \left(\frac{7\Delta N_{\rm eff}}{45N^2}\left(\frac{4}{11}\right)^{4 / 3}\right)^{1/4} T_{\gamma}^0,
\end{equation}
where $\Delta N_{\rm eff}=(N_{\mathrm{eff},95\%}-N_{\rm eff,SM})=0.23$~\cite{Aghanim:2018eyx}. Consequently,
\begin{equation}
T_{\rm CFT}^0\leq 0.034~{\rm meV}\frac{T_\gamma^0}{0.23~{\rm meV}} \left(\frac{N}{4.5}\right)^{-1/2} \,.
\end{equation}
As we have mentioned earlier, our model will be able to crunch away larger CC's if the CFT temperature is larger, hence we will always choose the maximal value of $T_{\rm CFT}^0$ consistent with the above bound.

\subsection{Uncertainties and Caveats of the Calculation}
\label{app:uncertainties}

We have encountered several parameters and expressions which cannot be calculated using our simple approach. These parameters affect (to varying degrees) the value of the maximal cosmological constant that can be crunched away. Here we will list each source of uncertainty, how we have estimated them,  and briefly discuss their effects.  
\begin{itemize}
	\item{$\alpha(T)$, the coefficient in Eq.~\eqref{eq:friction} determining  the energy lost to friction after exiting the region of validity in the O(3) bounce calculation. For the lower bound estimate we take $\alpha=0$, and for the numerical estimate with $\chi\left(r_{\rm max}\right) \to 0$ the calculation is continued into the non-calculable regime which implies an effective $\alpha(T)$ of order  ${\cal O}(1)$ for $\epsilon_2 \sim {\cal O}(1)$.}
	\item{$\Gamma_0^{(3)}$, the coefficient for the rate of bubble nucleation in Eq.~\eqref{eq:decayrate} due to the $O(3)$-symmetric action. There is an unknown functional determinant of dimension-3 that multiplies the nucleation rate. As discussed in Ref.~\cite{Linde:1981zj}, we estimate this functional determinant to be the cube of a physical scale which we assumed to be $\mathcal{O}(0.1-10)T$, at $T>T_*$. This is expressed in Eq.~\eqref{eq:decayrate} as an uncertainty of $\Gamma_0^{(3)}\sim 10^{-3}-10^3$, and we use as the central value $\Gamma_0^{(3)}=1$. As discussed in App.~\ref{app:bouncetoday}, even for
		$\Gamma_0^{(3)}$ as large as $10^3$,   $S_3(T^0_{\rm CFT})/T^0_{\rm CFT}=280$ is a sufficient condition that our patch has not yet decayed.}
	\item{$\Gamma_0^{(4)}$,  the coefficient for the rate of bubble nucleation in Eq.~\eqref{eq:O4decayrate}  for the $O(4)$-symmetric action. In nearly identical fashion to the discussion of $\Gamma_0^{(3)}$, a functional determinant of dimension-4 multiplies the nucleation rate of O(4)-symmetric solution. The only dimensionful scale at zero temperature is $\chi_*$. Therefore it is reasonable to assume that the functional determinant is $\chi_*^4$ up to a factor of $(\mathcal{O}(0.1-10))^4$. This results in the range $\Gamma_0^{(4)}=10^{-4}-10^4$. }
	\item{$S_3^{\rm non-calc}$, the action for the region where the dilaton description breaks down. For the lower bound calculation we have taken $S_3^{\rm non-calc}=0$.  This a conservative assumption becuase $S_3^{\rm non-calc}$ is typically positive, as one can easily see in a simple case without troughs and peaks. It should be noted though that no general proof for the positivity of this contribution exists. The numerical estimate with $\chi\left(r_{\rm max}\right) \to 0$, effectively includes a contribution to $S_3^{\rm non-calc}$, which is ${\cal O}(N^2)$,  for $\epsilon_2 \sim {\cal O}(1)$.}	
	\item{$S_4$, the bounce action for O(4) symmetric bubbles. Our estimation $S_4\sim 3N^2/16$, is purely based on dimensional analysis, and in fact we do not model the potential at zero temperature at all for the entire range of the $S_4$ calculation. We therefore vary $S_4$ between $3N^2/32$ and $3N^2/8$. Our $\Lambda_{\rm max}\propto e^{-S_4/8}$ is exponentially sensitive to $S_4$. Therefore, among the parameters discussed here, $S_4$ is the one to which we are most sensitive.}
\end{itemize}

\bibliographystyle{JHEP}
\bibliography{CC.bib}

\providecommand{\href}[2]{#2}\begingroup\raggedright\begin{thebibliography}{10}

\bibitem{Weinberg:1988cp}
S.~Weinberg, {\it {The Cosmological Constant Problem}},  {\em Rev. Mod. Phys.}
  {\bf 61} (1989) 1--23.

\bibitem{Martin:2012bt}
J.~Martin, {\it {Everything You Always Wanted to Know About the Cosmological
  Constant Problem (But Were Afraid to Ask)}},  {\em Comptes Rendus Physique}
  {\bf 13} (2012) 566--665, [\href{http://arxiv.org/abs/1205.3365}{{\tt
  arXiv:1205.3365}}].

\bibitem{Zeldovich:1968ehl}
{\relax Ya}.~B. Zel'dovich, A.~Krasinski, and {\relax Ya}.~B. Zeldovich, {\it
  {The Cosmological constant and the theory of elementary particles}},  {\em
  Sov. Phys. Usp.} {\bf 11} (1968) 381--393.

\bibitem{Fitch:1997cf}
{\it {P.J. Steinhardt, in Critical Problems in Physics, edited by V.L. Fitch
  and Dr.R. Marlow (Princeton University Press, Princeton, N. J., 1997).}}, .

\bibitem{Nobbenhuis:2004wn}
S.~Nobbenhuis, {\it {Categorizing Different Approaches to the Cosmological
  Constant Problem}},  {\em Found. Phys.} {\bf 36} (2006) 613--680,
  [\href{http://arxiv.org/abs/gr-qc/0411093}{{\tt gr-qc/0411093}}].

\bibitem{Polchinski:2006gy}
J.~Polchinski, {\it {The Cosmological Constant and the String Landscape}},  in
  {\em {The Quantum Structure of Space and Time: Proceedings of the 23Rd Solvay
  Conference on Physics. Brussels, Belgium. 1 - 3 December 2005}},
  pp.~216--236, 2006.
\newblock \href{http://arxiv.org/abs/hep-th/0603249}{{\tt hep-th/0603249}}.

\bibitem{Bousso:2007gp}
R.~Bousso, {\it {TASI Lectures on the Cosmological Constant}},  {\em Gen. Rel.
  Grav.} {\bf 40} (2008) 607--637, [\href{http://arxiv.org/abs/0708.4231}{{\tt
  arXiv:0708.4231}}].

\bibitem{Yoo2012}
J.~Yoo and Y.~Watanabe, {\it {Theoretical Models of Dark Energy}},  {\em Int.
  J. Mod. Phys.} {\bf D21} (2012) 1230002,
  [\href{http://arxiv.org/abs/1212.4726}{{\tt arXiv:1212.4726}}].

\bibitem{Padilla2015}
A.~Padilla, {\it {Lectures on the Cosmological Constant Problem}},
  \href{http://arxiv.org/abs/1502.05296}{{\tt arXiv:1502.05296}}.

\bibitem{Copeland2006}
E.~J. Copeland, M.~Sami, and S.~Tsujikawa, {\it {Dynamics of dark energy}},
  {\em Int. J. Mod. Phys.} {\bf D15} (2006) 1753--1936,
  [\href{http://arxiv.org/abs/hep-th/0603057}{{\tt hep-th/0603057}}].

\bibitem{Sola:2013gha}
J.~Sola, {\it {Cosmological constant and vacuum energy: old and new ideas}},
  {\em J. Phys. Conf. Ser.} {\bf 453} (2013) 012015,
  [\href{http://arxiv.org/abs/1306.1527}{{\tt arXiv:1306.1527}}].

\bibitem{Zumino:1974bg}
B.~Zumino, {\it {Supersymmetry and the Vacuum}},  {\em Nucl. Phys.} {\bf B89}
  (1975) 535.

\bibitem{Bellazzini:2013fga}
B.~Bellazzini, C.~Cs\'aki, J.~Hubisz, J.~Serra, and J.~Terning, {\it {A
  Naturally Light Dilaton and a Small Cosmological Constant}},  {\em Eur. Phys.
  J.} {\bf C74} (2014) 2790, [\href{http://arxiv.org/abs/1305.3919}{{\tt
  arXiv:1305.3919}}].

\bibitem{Coradeschi:2013gda}
F.~Coradeschi, P.~Lodone, D.~Pappadopulo, R.~Rattazzi, and L.~Vitale, {\it {A
  naturally light dilaton}},  {\em JHEP} {\bf 11} (2013) 057,
  [\href{http://arxiv.org/abs/1306.4601}{{\tt arXiv:1306.4601}}].

\bibitem{Witten:1995rz}
E.~Witten, {\it {Strong coupling and the cosmological constant}},  {\em Mod.
  Phys. Lett.} {\bf A10} (1995) 2153--2156,
  [\href{http://arxiv.org/abs/hep-th/9506101}{{\tt hep-th/9506101}}].

\bibitem{Abbott:1984qf}
L.~F. Abbott, {\it {A Mechanism for Reducing the Value of the Cosmological
  Constant}},  {\em Phys. Lett.} {\bf 150B} (1985) 427--430.

\bibitem{Steinhardt:2006bf}
P.~J. Steinhardt and N.~Turok, {\it {Why the cosmological constant is small and
  positive}},  {\em Science} {\bf 312} (2006) 1180--1182,
  [\href{http://arxiv.org/abs/astro-ph/0605173}{{\tt astro-ph/0605173}}].

\bibitem{ArkaniHamed:2002fu}
N.~Arkani-Hamed, S.~Dimopoulos, G.~Dvali, and G.~Gabadadze, {\it {Nonlocal
  Modification of Gravity and the Cosmological Constant Problem}},
  \href{http://arxiv.org/abs/hep-th/0209227}{{\tt hep-th/0209227}}.

\bibitem{Banks:2018jqo}
T.~Banks and W.~Fischler, {\it {Why the Cosmological Constant is a Boundary
  Condition}},  \href{http://arxiv.org/abs/1811.00130}{{\tt arXiv:1811.00130}}.

\bibitem{Sundrum:1997js}
R.~Sundrum, {\it {Towards an effective particle string resolution of the
  cosmological constant problem}},  {\em JHEP} {\bf 07} (1999) 001,
  [\href{http://arxiv.org/abs/hep-ph/9708329}{{\tt hep-ph/9708329}}].

\bibitem{Kaloper:2013zca}
N.~Kaloper and A.~Padilla, {\it {Sequestering the Standard Model Vacuum
  Energy}},  {\em Phys. Rev. Lett.} {\bf 112} (2014), no.~9 091304,
  [\href{http://arxiv.org/abs/1309.6562}{{\tt arXiv:1309.6562}}].

\bibitem{Anderson:1971pn}
J.~L. Anderson and D.~Finkelstein, {\it {Cosmological constant and fundamental
  length}},  {\em Am. J. Phys.} {\bf 39} (1971) 901--904.

\bibitem{Linde:1988ws}
A.~D. Linde, {\it {The Universe Multiplication and the Cosmological Constant
  Problem}},  {\em Phys. Lett.} {\bf B200} (1988) 272--274.

\bibitem{Weinberg:1987dv}
S.~Weinberg, {\it {Anthropic Bound on the Cosmological Constant}},  {\em Phys.
  Rev. Lett.} {\bf 59} (1987) 2607.

\bibitem{Vilenkin:1983xq}
A.~Vilenkin, {\it {The Birth of Inflationary Universes}},  {\em Phys. Rev.}
  {\bf D27} (1983) 2848.

\bibitem{Gibbons:1984hx}
P.~Steinhardt, {\it {in The Very Early Universe. Proceedings, Nuffield
  Workshop, Cambridge, UK, eds. G.~Gibbons, S.~Hawking, and S.~Siklos}}, .

\bibitem{Linde:1993xx}
A.~D. Linde, D.~A. Linde, and A.~Mezhlumian, {\it {From the Big Bang Theory to
  the Theory of a Stationary Universe}},  {\em Phys. Rev.} {\bf D49} (1994)
  1783--1826, [\href{http://arxiv.org/abs/gr-qc/9306035}{{\tt gr-qc/9306035}}].

\bibitem{Linde:2010xz}
A.~Linde and M.~Noorbala, {\it {Measure Problem for Eternal and Non-Eternal
  Inflation}},  {\em JCAP} {\bf 1009} (2010) 008,
  [\href{http://arxiv.org/abs/1006.2170}{{\tt arXiv:1006.2170}}].

\bibitem{Freivogel2011}
B.~Freivogel, {\it {Making predictions in the multiverse}},  {\em Class. Quant.
  Grav.} {\bf 28} (2011) 204007, [\href{http://arxiv.org/abs/1105.0244}{{\tt
  arXiv:1105.0244}}].

\bibitem{Coleman:1980aw}
S.~R. Coleman and F.~De~Luccia, {\it {Gravitational Effects on and of Vacuum
  Decay}},  {\em Phys. Rev.} {\bf D21} (1980) 3305.

\bibitem{Aghanim:2018eyx}
{\bf Planck} Collaboration, N.~Aghanim et~al., {\it {Planck 2018 Results. Vi.
  Cosmological Parameters}},  \href{http://arxiv.org/abs/1807.06209}{{\tt
  arXiv:1807.06209}}.

\bibitem{Barbieri:2000gf}
R.~Barbieri and A.~Strumia, {\it {The `Lep Paradox'}},  in {\em {4Th Rencontres
  Du Vietnam: Physics at Extreme Energies (Particle Physics and Astrophysics)
  Hanoi, Vietnam, July 19-25, 2000}}, 2000.
\newblock \href{http://arxiv.org/abs/hep-ph/0007265}{{\tt hep-ph/0007265}}.

\bibitem{Starobinsky:1980te}
A.~A. Starobinsky, {\it {A New Type of Isotropic Cosmological Models without
  Singularity}},  {\em Phys. Lett.} {\bf 91B} (1980) 99--102. [,771(1980)].

\bibitem{Guth:1980zm}
A.~H. Guth, {\it {The Inflationary Universe: a Possible Solution to the Horizon
  and Flatness Problems}},  {\em Phys. Rev.} {\bf D23} (1981) 347--356. [Adv.
  Ser. Astrophys. Cosmol.3,139(1987)].

\bibitem{Linde:1981mu}
A.~D. Linde, {\it {A New Inflationary Universe Scenario: a Possible Solution of
  the Horizon, Flatness, Homogeneity, Isotropy and Primordial Monopole
  Problems}},  {\em Phys. Lett.} {\bf 108B} (1982) 389--393. [Adv. Ser.
  Astrophys. Cosmol.3,149(1987)].

\bibitem{Martin:2013tda}
J.~Martin, C.~Ringeval, and V.~Vennin, {\it {Encyclopædia Inflationaris}},
  {\em Phys. Dark Univ.} {\bf 5-6} (2014) 75--235,
  [\href{http://arxiv.org/abs/1303.3787}{{\tt arXiv:1303.3787}}].

\bibitem{Brown:1987dd}
J.~D. Brown and C.~Teitelboim, {\it {Dynamical Neutralization of the
  Cosmological Constant}},  {\em Phys. Lett.} {\bf B195} (1987) 177--182.

\bibitem{Brown:1988kg}
J.~D. Brown and C.~Teitelboim, {\it {Neutralization of the Cosmological
  Constant by Membrane Creation}},  {\em Nucl. Phys.} {\bf B297} (1988)
  787--836.

\bibitem{Arvanitaki:2009fg}
A.~Arvanitaki, S.~Dimopoulos, S.~Dubovsky, N.~Kaloper, and J.~March-Russell,
  {\it {String Axiverse}},  {\em Phys. Rev.} {\bf D81} (2010) 123530,
  [\href{http://arxiv.org/abs/0905.4720}{{\tt arXiv:0905.4720}}].

\bibitem{Bousso:2000xa}
R.~Bousso and J.~Polchinski, {\it {Quantization of four form fluxes and
  dynamical neutralization of the cosmological constant}},  {\em JHEP} {\bf 06}
  (2000) 006, [\href{http://arxiv.org/abs/hep-th/0004134}{{\tt
  hep-th/0004134}}].

\bibitem{Bachlechner:2017zpb}
T.~C. Bachlechner, K.~Eckerle, O.~Janssen, and M.~Kleban, {\it {Multiple-Axion
  Framework}},  {\em Phys. Rev.} {\bf D98} (2018), no.~6 061301,
  [\href{http://arxiv.org/abs/1703.00453}{{\tt arXiv:1703.00453}}].

\bibitem{Witten:1998zw}
E.~Witten, {\it {Anti-de~Sitter Space, Thermal Phase Transition, and
  Confinement in Gauge Theories}},  {\em Adv. Theor. Math. Phys.} {\bf 2}
  (1998) 505--532, [\href{http://arxiv.org/abs/hep-th/9803131}{{\tt
  hep-th/9803131}}].

\bibitem{Creminelli:2001th}
P.~Creminelli, A.~Nicolis, and R.~Rattazzi, {\it {Holography and the
  Electroweak Phase Transition}},  {\em JHEP} {\bf 03} (2002) 051,
  [\href{http://arxiv.org/abs/hep-th/0107141}{{\tt hep-th/0107141}}].

\bibitem{Callan:1977pt}
C.~G. Callan, Jr. and S.~R. Coleman, {\it {The Fate of the False Vacuum. 2.
  First Quantum Corrections}},  {\em Phys. Rev.} {\bf D16} (1977) 1762--1768.

\bibitem{Coleman:1977py}
S.~R. Coleman, {\it {The Fate of the False Vacuum. 1. Semiclassical Theory}},
  {\em Phys. Rev.} {\bf D15} (1977) 2929--2936. [Erratum: Phys.
  Rev.D16,1248(1977)].

\bibitem{Linde:1981zj}
A.~D. Linde, {\it {Decay of the False Vacuum at Finite Temperature}},  {\em
  Nucl. Phys.} {\bf B216} (1983) 421. [Erratum: Nucl. Phys.B223,544(1983)].

\bibitem{Randall:1999ee}
L.~Randall and R.~Sundrum, {\it {A Large Mass Hierarchy from a Small Extra
  Dimension}},  {\em Phys. Rev. Lett.} {\bf 83} (1999) 3370--3373,
  [\href{http://arxiv.org/abs/hep-ph/9905221}{{\tt hep-ph/9905221}}].

\bibitem{vonHarling:2017yew}
B.~von Harling and G.~Servant, {\it {QCD-induced Electroweak Phase
  Transition}},  {\em JHEP} {\bf 01} (2018) 159,
  [\href{http://arxiv.org/abs/1711.11554}{{\tt arXiv:1711.11554}}].

\bibitem{Baratella:2018pxi}
P.~Baratella, A.~Pomarol, and F.~Rompineve, {\it {The Supercooled Universe}},
  {\em JHEP} {\bf 03} (2019) 100, [\href{http://arxiv.org/abs/1812.06996}{{\tt
  arXiv:1812.06996}}].

\bibitem{Goldberger:1999uk}
W.~D. Goldberger and M.~B. Wise, {\it {Modulus Stabilization with Bulk
  Fields}},  {\em Phys. Rev. Lett.} {\bf 83} (1999) 4922--4925,
  [\href{http://arxiv.org/abs/hep-ph/9907447}{{\tt hep-ph/9907447}}].

\bibitem{Csaki:2004ay}
C.~Cs\'aki, {\it {TASI Lectures on Extra Dimensions and Branes}},  in {\em
  {From Fields to Strings: Circumnavigating Theoretical Physics. Ian Kogan
  Memorial Collection (3 Volume Set)}}, pp.~605--698, 2004.
\newblock \href{http://arxiv.org/abs/hep-ph/0404096}{{\tt hep-ph/0404096}}.

\bibitem{Hawking:1982dh}
S.~W. Hawking and D.~N. Page, {\it {Thermodynamics of Black Holes in anti-De
  Sitter Space}},  {\em Commun. Math. Phys.} {\bf 87} (1983) 577.

\bibitem{Agashe:2007zd}
K.~Agashe, H.~Davoudiasl, G.~Perez, and A.~Soni, {\it {Warped Gravitons at the
  Lhc and Beyond}},  {\em Phys. Rev.} {\bf D76} (2007) 036006,
  [\href{http://arxiv.org/abs/hep-ph/0701186}{{\tt hep-ph/0701186}}].

\bibitem{Mukhanov:2014uwa}
V.~Mukhanov, {\it {Inflation without Selfreproduction}},  {\em Fortsch. Phys.}
  {\bf 63} (2015) 36--41, [\href{http://arxiv.org/abs/1409.2335}{{\tt
  arXiv:1409.2335}}].

\bibitem{Csaki:1999mp}
C.~Cs\'aki, M.~Graesser, L.~Randall, and J.~Terning, {\it {Cosmology of brane
  models with radion stabilization}},  {\em Phys. Rev.} {\bf D62} (2000)
  045015, [\href{http://arxiv.org/abs/hep-ph/9911406}{{\tt hep-ph/9911406}}].

\bibitem{Csaki:2000zn}
C.~Cs\'aki, M.~L. Graesser, and G.~D. Kribs, {\it {Radion dynamics and
  electroweak physics}},  {\em Phys. Rev.} {\bf D63} (2001) 065002,
  [\href{http://arxiv.org/abs/hep-th/0008151}{{\tt hep-th/0008151}}].

\bibitem{Rattazzi:2000hs}
R.~Rattazzi and A.~Zaffaroni, {\it {Comments on the holographic picture of the
  Randall-Sundrum model}},  {\em JHEP} {\bf 04} (2001) 021,
  [\href{http://arxiv.org/abs/hep-th/0012248}{{\tt hep-th/0012248}}].

\bibitem{ArkaniHamed:2000ds}
N.~Arkani-Hamed, M.~Porrati, and L.~Randall, {\it {Holography and
  phenomenology}},  {\em JHEP} {\bf 08} (2001) 017,
  [\href{http://arxiv.org/abs/hep-th/0012148}{{\tt hep-th/0012148}}].

\bibitem{Csaki:2007ns}
C.~Cs\'aki, J.~Hubisz, and S.~J. Lee, {\it {Radion phenomenology in realistic
  warped space models}},  {\em Phys. Rev.} {\bf D76} (2007) 125015,
  [\href{http://arxiv.org/abs/0705.3844}{{\tt arXiv:0705.3844}}].

\end{thebibliography}\endgroup

\end{document}